\documentclass[journal,twocolumn]{IEEEtran}

\ifCLASSINFOpdf
\else
\fi
\usepackage{cite, amsfonts, amssymb, amsthm, bm, bbm, graphicx, relsize, multirow, booktabs, tikz,subfigure,soul}
\usepackage{ucs}
\usepackage[utf8x]{inputenc}
\usepackage[cmex10]{amsmath}
\usepackage{cite, amsfonts, amssymb, amsthm, bm, bbm, graphicx, relsize, multirow, booktabs, tikz,subfigure,soul}
\usepackage[american]{babel}
\usepackage[T1]{fontenc}
\usepackage{algorithmic, algorithm}
\usepackage[multiple]{footmisc}
\theoremstyle{definition}

\theoremstyle{remark}

\newcommand{\ds}{\displaystyle}

\begin{document}
This paper has been accepted for publication on the IEEE Transactions on Communications

\copyright 2019 IEEE. Personal use of this material is permitted. Permission from IEEE must be obtained for all other uses, in any current or future media, including reprinting/republishing this material for advertising or promotional purposes, creating new collective works, for resale or redistribution to servers or lists, or reuse of any copyrighted  component of this work in other works.”

\newpage

\title{MIMO-UFMC Transceiver Schemes for Millimeter Wave Wireless Communications}

\author{Stefano Buzzi, {\em Senior Member}, {\em IEEE},  Carmen D'Andrea,
{\em Student Member}, {\em IEEE}, Dejian Li, and Shulan Feng
\thanks{This paper has been supported by Huawei through HIRP OPEN Agreement No. HO2016050002BM.}
\thanks{This paper has been partly presented at the \textit{2018 IEEE 29th Annual International Symposium on Personal, Indoor, and Mobile Radio Communications (PIMRC), Workshop on Millimeter Waves Communications},  Bologna, Italy, September 2018 and at \textit{Chinacom 2018 - 13th EAI International Conference on Communications and Networking in China}, Chengdu, China, October 2018. 
}
\thanks{S. Buzzi and C. D'Andrea  are with the Department of Electrical and Information Engineering, University of Cassino and Lazio Meridionale, I-03043 Cassino, Italy (\{buzzi, carmen.dandrea\}@unicas.it); they are also affiliated with Consorzio Nazionale Interuniversitario per le Telecomunicazioni (CNIT), Italy.
D. Li and S. Feng are with Hisilicon Technologies Co., Ltd, Beijing, P. R. China 
(\{lidejian, shulan.feng\}@hisilicon.com).
}
}
\maketitle

%



\maketitle

\begin{abstract}
The UFMC modulation is among the most considered solutions for the realization of beyond-OFDM air interfaces for future wireless 
networks. This paper focuses on the design and analysis of an UFMC transceiver equipped with multiple antennas and operating at millimeter wave carrier frequencies. The paper provides the full mathematical model of a MIMO-UFMC transceiver, taking into account the presence of hybrid analog/digital beamformers at both ends of the communication links. Then, several detection structures are proposed, both for the case of single-packet isolated transmission, and for the case of multiple-packet continuous transmission. In the latter situation, the paper also considers the case in which no guard time among adjacent packets is inserted, trading off an increased level of interference with higher values of spectral efficiency. At the analysis stage, the several considered detection structures and transmission schemes are compared in terms of bit-error-rate, root-mean-square-error, and system throughput. The numerical results show that the proposed transceiver algorithms are effective and that the linear MMSE data detector is capable of well managing the increased interference brought by the removal of guard times among consecutive packets, thus yielding throughput gains of about 10 - 13 $\%$. The effect of phase noise at the receiver is also numerically assessed, and it is shown that the recursive implementation of the linear MMSE exhibits some degree of robustness against this disturbance. 
\end{abstract}


%
\IEEEpeerreviewmaketitle

\section{Introduction}
The research activity on modulation formats alternative to orthogonal frequency division multiplexing (OFDM) for the 
fifth-generation (5G) and beyond of wireless cellular systems has been very intensive in the last decade \cite{BaBuCoMoRuUg14,Farhang-Boroujeny201192,buzzi2016introduction}. 
Indeed, even though the first 5G New Radio  standalone specification still relies on OFDM with flexible numerology\cite{parkvall_2018}, 3GPP has not yet thoroughly addressed the use-case of massive machine-type-communications (mMTC) as well as the choice of the  modulation format at above-30 GHz carrier frequencies; moreover, new use-cases are recently arising, such as high data-rate mMTC, which indicate that the dispute about the modulation formats and waveforms is not at all close to its end. 
Based on the consideration that OFDM exhibits some key drawbacks such as considerable out-of-band (OOB) emissions, spectral inefficiency due to the use of the cyclic prefix, vulnerability to non-linearities in  power amplifiers due to large peak-to-average-power-ratio (PAPR), and considerable inter-carrier interference in the case of imperfect timing and frequency synchronization, several alternative modulation schemes have been proposed, such as filter bank multicarrier (FBMC) \cite{farhang2014filter,Farhang-Boroujeny201192,nissel2017filter}, filtered-OFDM 
\cite{filtered_OFDM2015}, flexible-OFDM \cite{cai2012flexible}, 
 weighted overlap and add
based OFDM (WOLA-OFDM) \cite{zayani2016wola}
generalized frequency division multiplexing (GFDM) \cite{Fettweis2009,michailow2014generalized}, 
universal filtered multicarrier (UFMC) \cite{Vakilian_UFMC2013,schaich2014waveform2}, index modulation \cite{basar2016index}.

Among these, UFMC modulation  has received prominent attention. Introduced in \cite{wunder2013system} in the framework of the EU-funded Horizon2020 research project 5G-NOW, UFMC is an intermediate scheme between filtered-OFDM and FBMC. While in filtered OFDM the whole OFDM signal is filtered to reduce OOB emissions and achieve better spectral containment\cite{filtered_OFDM2015}, and while in FBMC each subcarrier is individually filtered \cite{Farhang-Boroujeny201192}, in UFMC the subcarriers are grouped in contiguous, non-overlapping blocks, called \textit{subbands}, and each subband is individually filtered \cite{Vakilian_UFMC2013}. Filtering at the subband level is motivated by the fact that time-frequency misalignment typically occurs between entire blocks of subcarriers; moreover, and mostly important, the adoption of subband-wise filtering permits employing filters with a larger bandwidth and, consequently, with a shorter impulse response than that of the filters used in FBMC.

First papers dealing with the UFMC modulation and highlighting its advantages with respect to OFDM were \cite{wunder2013system,5GNow-D3.1-2013,Vakilian_UFMC2013}. The paper \cite{schaich2014relaxed} 
shows that UFMC has increased robustness to timing and frequency synchronization errors and introduces the concept of autonomous timing advance,  a mechanism enabling the system to operate based on open-loop synchronization only. Papers \cite{mukherjee2015reduced,wen2018design,wang2015filter} deal with the problem of filter shape optimization; in particular, reference \cite{mukherjee2015reduced} applies Bohman filter-based pulse shaping with combination of antipodal symbol-pairs to the edge-subcarriers of
the subbands in order to reduce the spectral leakage into adjacent subbands; the paper \cite{wen2018design} designs nearly equi-ripple filters in the stopband by solving an optimization problem where the passband ripple is constrained and the maximum ripple in the stopband is minimized; in \cite{wang2015filter}, instead, a filter design procedure is proposed in order to minimize the leakage due to frequency and timing synchronization errors. In order to reduce interference among signals in adjacent subbands, papers
 \cite{wang2015universal,zhang2017universal} propose active interference cancellation schemes based on the solution of optimization problems aimed at determining the weighting coefficients for interference removal. The problem of channel estimation in UFMC transceivers is tackled in \cite{wang2015pilot}, while in 
 \cite{zhang2017subband} authors propose a "flexible" version of UFMC that permits integrating multiple frame structures with different subcarrier spacing in one radio carrier. 
  
Surprisingly enough, the above papers mainly deal with discrete-time single-antenna transceivers operating over a simple channel, usually modeled as a discrete-time linear-time-invariant filter. 
 On the other hand, it is well known that current (and future) wireless networks heavily rely on the use of multiple antennas, that are needed both to provide diversity performance gains and to multiplex several users on the same time-frequency resource slot. How UFMC modulation can be adapted to and take advantage of transceivers using multiple antennas is a topic that, to the best of authors' knowledge, appears to have been neglected so far.  Likewise, the above cited papers refer to the case in which sub-6 GHz carrier frequencies are used, while no studies are available on the use of UFMC at millimeter wave (mmWave) carrier frequencies. The use of mmWave, indeed, is one of the three key technologies\footnote{The other two technologies are the densification of the network with small-size cells and the use of large scale antenna arrays, i.e. massive MIMO.} needed to achieve the envisioned 1000x throughput gain with respect to the current generation of wireless networks \cite{whatwill5Gbe}. Using high carrier frequencies permits taking advantage of large and unused frequency bands, making it possible to seamlessly provide Gbit/s data rates 
to mobile users. On the other hand, mmWave carrier frequencies are essentially a short range (up to 100-200 m) technology given the increased path-loss and blockage effects \cite{6515173}, and \textit{need} to be used in conjunction with multiple antennas, since large array gains are useful to compensate for increased path-loss. Papers \cite{buzzi2017massive,bjornson2018massive} illustrate the main differences and technological challenges that the use of (possibly massive) MIMO transceivers poses at mmWave carrier frequencies with respect to traditional sub-6 GHz frequencies. In particular, at mmWave frequencies the channel impulse response has a sparse nature; hardware constraints prevent the adoption of fully digital beamforming structures \cite{han2015large}; and, moreover, phase noise, which
arises predominantly due to imperfections of the local oscillator (LO) in the transceivers, 
  has a detrimental effect on the system performance and cannot be neglected \cite{khanzadi2015phase}. Accordingly, the design of wireless transceivers for mmWave carrier frequencies has to face different challenges and constraints with respect to the ones that hold for systems at sub-6 GHz frequencies.

\subsection{Paper contribution}
This paper, to the best of authors' knowledge, is the first one to propose and analyze UFMC 
modulation for MIMO wireless link operating at mmWave carrier frequencies. The contribution of this paper may be summarized as follows. First of all, the paper provides the full mathematical model for a MIMO UFMC transceiver, taking into account the presence of hybrid analog/digital beamformers, and the use of a number of RF chains smaller than the number of antennas. 
Then, the paper proposes and analyzes several data detection structures, including linear minimum mean square error (MMSE) detectors, with different degrees of adaptiveness and tracking capabilities,  with different amounts of complexity, and that do not require an explicit channel estimation phase at the receiver. 
The proposed receivers are shown to exhibit much better performance than the traditional ones, as detailed for instance in 
\cite{Vakilian_UFMC2013}, especially when inter-packet interference is considered.
While traditionally UFMC packets are spaced by some guard-time, the paper also proposes a new "compact" UFMC transmission, wherein no guard interval is included among consecutive packets; interestingly, the MMSE-based receivers are capable to automatically adapt to the new transmission format and to combat the increased interference that it generates. In order to avoid the need for feeding back channel state information at the transmitter, the paper also proposes and analyzes the use of a novel channel-independent beamformer at the transmitter. Finally, the proposed data detection algorithms are also tested in the presence of phase noise  at the receiver.  Numerical results will show that the new "compact" UFMC format coupled with the MMSE receivers has the capability of providing increased throughput, as well as that the recursive implementations of the linear MMSE receiver exhibit increased robustness against the effects of receiver phase noise.

\medskip
This paper is organized as follows. Next section contains some preliminary derivations on the mathematical model of a single-antenna UFMC transceiver and on the MIMO channel model at mmWave  carrier frequencies. 
Section III shows how UFMC can be coupled with the use of multiple antennas, by providing also some transceiver algorithms for the case of single-packet transmission. In Section IV we consider the case in which a linear MMSE equalizer and data detector is used at the receiver; several versions of this structure are proposed, one based on the use of time-averaged batch estimates, and two others based on the use of recursive algorithms with learning capabilities. The use of a channel independent beamformer is also illustrated in Section IV. Section V deals with the system performance analysis, and thus contains the definition of the used performance measures,
a theoretical analysis about the system bit-error-rate (BER), and the discussion about the obtained numerical results. Finally, concluding remarks are given in Section VI, while briefs on the used model for the phase noise are reported in the Appendix. 

\subsection{Notation}
\noindent
The following notation is used in the paper. Bold lowercase letters (such as $\mathbf{a}$) denote column vectors, bold uppercase letters (such as $\mathbf{A}$) denote matrices, non-bold letters $a$ and $A$ denote scalar values. The transpose, the inverse, the Moore-Penrose generalized inverse and the conjugate transpose of a matrix $\mathbf{A}$ are denoted by $\mathbf{A}^T$, $\mathbf{A}^{-1}$, $\mathbf{A}^{+}$ and $\mathbf{A}^H$, respectively. The trace of the matrix $\mathbf{A}$ is denoted as tr$\left(\mathbf{A}\right)$. The $n$-th entry of the vector $\mathbf{a}$ is denoted as $\mathbf{a}_{(n)}$ and the $n$-th column and the $n$-th row of the matrix $\mathbf{A}$ are denoted as $\mathbf{A}(:,n)$ and $\mathbf{A}(n,:)$, respectively. The $N$-dimensional identity matrix is denoted as $\mathbf{I}_N$ and the $(N \times M)$-dimensional matrix with all zero entries is denoted as $\mathbf{0}_{N \times M}$. The vectorization operator is denoted by vec$(\cdot)$ and the Kronecker product is denoted by $\otimes$. The block-diagonal matrix obtained from matrices $\mathbf{A}_1, \ldots, \mathbf{A}_N$ is denoted by blkdiag$\left( \mathbf{A}_1, \ldots, \mathbf{A}_N\right)$. The  Dirac's delta pulse is denoted as $\delta(t)$. The statistical expectation operator is denoted as $\mathbb{E}[\cdot]$; $\mathcal{CN}\left(\mu,\sigma^2\right)$ denotes  a complex circularly symmetric Gaussian random variable with mean $\mu$ and variance $\sigma^2$, while $\mathcal{U}(a,b)$ denotes a random variable that is uniformly distributed in $[a,b]$. The complementary error function is denoted as erfc$(\cdot)$.

\section{Preliminaries}
\subsection{Review of single-antenna UFMC modulation}
We start with providing the mathematical model of the UFMC modulation format in a single-antenna scenario. 
We build upon papers \cite{schaich2014waveform2,chen2014multiple} and provide the mathematical description of the block-schemes therein reported. 
This model will be then used in Section III to design the MIMO-UFMC transceiver operating at mmWave frequencies. 

\begin{figure}[t]
\includegraphics[scale=0.22]{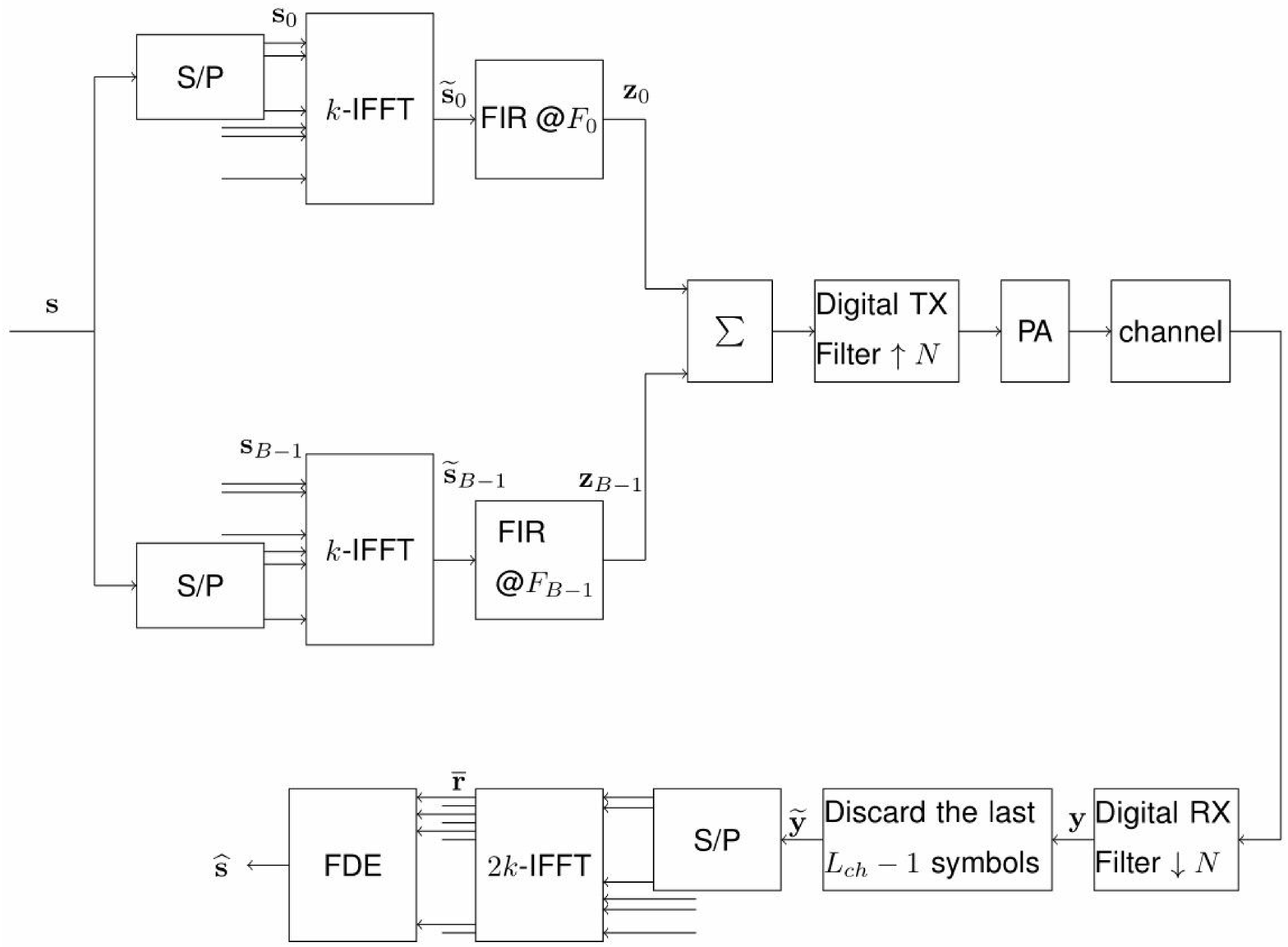}
\caption{Block scheme of the UFMC single antenna transceiver.}
\label{fig:UFMC_scheme_singleantenna}
\end{figure}

We refer to the block-scheme reported in Fig. \ref{fig:UFMC_scheme_singleantenna}. We assume that the $k$ subcarriers are split in $B$ subbands of $D$ subcarriers each (thus implying that $k=BD$). 
We focus on the transmission of a single packet  of $k$ data symbols, arranged into the $k$-dimensional vector $\mathbf{s}$;
the case of multiple packet   transmission will be addressed with reference to the general MIMO architecture in the next section.  Let $\mathbf{s}_i$ be a $k$-dimensional vector
whose $n$-th entry $\mathbf{s}_{i, \, (n)}$ is
 defined as follows
\begin{equation}
\mathbf{s}_{i ,\, (n)} \triangleq \left\{ \begin{array}{lll} \mathbf{s}_{(n)} & n=iD, iD+1, \ldots, (i+1)D-1 \\
0 & \mbox{otherwise} \end{array} \right. \; ,
\end{equation}
for $i=0, \ldots, B-1$ and $n=0, \ldots, k-1$. 
Defining the matrices 
$$
\mathbf{P}_i =\mbox{diag}\left([\; \underbrace{0 \ldots 0}_{iD} \underbrace{1 \ldots 1}_D \underbrace{0 \ldots 0}_{k-(i+1)D}]\right) \; , \quad i=0, \ldots, B-1 \; ,
$$
it is easily seen that $\mathbf{s}_i=\mathbf{P}_i \mathbf{s}$. The vectors $\mathbf{s}_i$ go through an IFFT transformation; letting $\mathbf{W}_{k, IFFT}$ denote the $(k \times k)$-dimensional matrix representing the isometric IFFT transformation\footnote{The $(m,n)$-th entry of $\mathbf{W}_{k, IFFT}$ is thus $\frac{1}{\sqrt{k}} e^{j 2 \pi (m-1)(n-1)/k}$.}, we have
\begin{equation}
\widetilde{\mathbf{s}}_i=\mbox{IFFT}(\mathbf{s}_i)=\mathbf{W}_{k,IFFT}\mathbf{P}_i\mathbf{s} \; .
\end{equation} 
The vectors $\widetilde{\mathbf{s}}_i$ then undergo a finite impulse response (FIR) passband filtering in order to improve their frequency localization property. Any passband FIR filter can be used; a customary choice is to resort to Dolph-Chebyshev discrete-time window, that permits controlling the side-lobes level with respect to the peak of the main lobe. 
Denoting by $\mathbf{g}\triangleq [g_0, g_1, \ldots, g_{L-1}]^T$ the $L$-dimensional vector representing the Dolph-Chebyshev prototype filter, the FIR filter used in the $i$-th subband to process the vector $\widetilde{\mathbf{s}}_i$ is denoted by $\mathbf{g}_i$ and its entries $g_{i,0}, g_{i,1}, \ldots, g_{i,L-1}$  are defined as follows
\begin{equation}
g_{i,\ell}=g_i e^{j 2 \pi \frac{F_i \ell}{k}} \; , i=0, \ldots, B-1, \; \; \ell=0, \ldots, L-1 \; ,
\end{equation}
with $F_i \triangleq \frac{D-1}{2} + iD$ the normalized frequency shift of the filter tuned to the $i$-th subband. 
Denoting by $\mathbf{G}_i$ the toeplitz $[(k+L-1) \times  k]$-dimensional matrix describing the discrete convolution operation with the filter $\mathbf{g}_i$, at the output we have the following  $(k+L-1)$-dimensional  vector
\begin{equation}
\mathbf{z}_i=\mathbf{G}_i \widetilde{\mathbf{s}}_i= \mathbf{G}_i \mathbf{W}_{k,IFFT}\mathbf{P}_i\mathbf{s} \; .
\end{equation}
The vectors $\mathbf{z}_i$ are summed together, multiplied by the amplification factor $\sqrt{P_T}$ and transmitted. The propagation channel is modeled as a discrete-time FIR filter of length $L_{ch}$, represented by the $L_{ch}$-dimensional vector $\mathbf{h}$. Denoting by $\mathbf{M}_h$ the $[(k+L+L_{ch}-2) \times (k+L-1)]$-dimensional toeplitz matrix describing the convolution with the channel $\mathbf{h}$, the discrete-time version of the received signal is written as the following $(k+L+L_{ch}-2)$-dimensional vector
\begin{equation}
\mathbf{y}=\sqrt{P_T}\mathbf{M}_h\left( \ds\sum_{i=0}^{B-1}\mathbf{G}_i \mathbf{W}_{k,IFFT}\mathbf{P}_i\right) \mathbf{s}+ \mathbf{w} \; ,
\label{eq:receivedUFMC}
\end{equation}
with $\mathbf{w}$ the additive white Gaussian noise contribution. Given Eq. \eqref{eq:receivedUFMC}, it is easily recognized that we have a standard linear model for the received signal, similar for instance to what happens in multiuser code-division multiple access systems or in single-user MIMO links,  and several linear and non-linear data detection schemes can be used. Under the considered ideal conditions\footnote{Note that we are considering the isolated transmission of a single UFMC block (i.e. no interblock interference is present), perfectly linear power amplifiers, no phase noise at the receiver, and perfect time-frequency synchronization.}, however, a very simple processing may be used based, similarly to what happens in OFDM, on the use of an FFT and of a simple one-tap frequency-domain equalization. In particular, first of all the last $L_{ch}-1$ symbols are removed from the received vector $\mathbf{y}$. Letting $\mathbf{D}_{L_{ch}-1}$ be a 
$[(k+L-1) \times (k+L+L_{ch}-2)]$-dimensional matrix defined as
\begin{equation}
\mathbf{D}_{L_{ch}-1}=\left[ \mathbf{I}_{k+L-1} \; \, \mathbf{0}_{k+L-1,L_{ch}-1}
\right]
\end{equation}
in matrix notation we have
\begin{equation}
\widetilde{\mathbf{y}}=\mathbf{D}_{L_{ch}-1} \mathbf{y} \; .
\label{eq:ytilde_UFMC}
\end{equation}
The vector $\widetilde{\mathbf{y}}$, of dimension $k+L-1$ is then zero-padded and FFTed on $2k$ points. Letting 
$\mathbf{W}_{2k,FFT}$ denote the $2k$-points isometric FFT\footnote{The $(m,n)$-th entry of $\mathbf{W}_{2k,FFT}$ is expressed as $\frac{1}{\sqrt{2k}}e^{-j 2 \pi (m-1)(n-1)/(2k)}$.}, we thus obtain the $2k$-dimensional vector 
$\mathbf{r}=\mathbf{W}_{2k,FFT} \mathbf{y}$. This vector is finally downsampled by a factor of 2 to obtain a $k$-dimensional vector, that we denote by $\overline{\mathbf{r}}$.
Now, under the cited ideal conditions, an almost interference-free soft estimate of the transmitted data vector $\mathbf{s}$ can be easily obtained. Indeed, letting 
$\widetilde{\mathbf{h}}_{(n)}$ and $\widetilde{\mathbf{g}}_{i,(n)}$  denote the $n$-th coefficient of the isometric $2k$-points FFT of the channel vector $\mathbf{h}$ and of the $i$-th subband filter $\mathbf{g}_i$, respectively, it can be shown that 
the $q$-th entry of $\overline{\mathbf{r}}$ is expressed as
\begin{equation}
\overline{\mathbf{r}}_{(q)} \approx \frac{2k}{\sqrt{2}} \widetilde{\mathbf{h}}_{(2q-1)} \mathcal{G}_{\lfloor q/D \rfloor}(2q-1) \mathbf{s}_{(q)} + \widetilde{\mathbf{w}}_{(2q-1)},
\label{eq:UFMC_approx}
\end{equation}
with $q=1, \ldots, k$, and 
with $ \widetilde{\mathbf{w}}_{(\ell)}$ the $\ell$-th coefficient of the isometric $2k$-points FFT of the noise vector $\mathbf{w}$. 
Notice that Eq. \eqref{eq:UFMC_approx} does not hold with perfect equality since we, according to the main UFMC references, have removed the last $L_{ch}-1$ symbols from the received vector $\mathbf{y}$. Actually, for the case of a single-packet isolated transmission there is no reason to remove these samples (provided that $L+L_{ch}-2 \leq k$), so that Eq. \eqref{eq:UFMC_approx} may hold with an equality sign.
It is thus seen that data symbols are almost interference-free, and so 
a soft estimate of the $q$-th entry of $\mathbf{s}$ is obtained as
\begin{equation}
\begin{array}{lll}
\ds \widehat{\mathbf{s}}_{(q)}& \approx \ds \frac{\sqrt{2}}{2k} \frac{\overline{\mathbf{r}}_{(q)}}{\widetilde{\mathbf{h}}_{(2q-1)} \widetilde{\mathbf{g}}_{\lfloor q/D \rfloor, (2q-1)}} \\ &  \ds=\mathbf{s}_{(q)} +  \frac{\sqrt{2}}{2k} \frac{\widetilde{\mathbf{w}}_{(2q-1)}}{\widetilde{\mathbf{h}}_{(2q-1)} \widetilde{\mathbf{g}}_{\lfloor q/D \rfloor, (2q-1)} }\; .
\label{eq:FDE_UFMC}
\end{array}
\end{equation}
 The processing reported in \eqref{eq:FDE_UFMC} represents the frequency-domain equalizer (FDE) block reported in Fig. \ref{fig:UFMC_scheme_singleantenna}.
 
\medskip 
\noindent
\textit{UFMC versus OFDM complexity comparison.} UFMC transceivers have a higher complexity than OFDM transceivers. Indeed, as seen from Fig \ref{fig:UFMC_scheme_singleantenna}, the UFMC  transmitter has $B$ (the number of subbands) $k$-IFFTs and $B$ filtering operations, while at the receiver  a $2k$-FFT is required. Conversely, in OFDM, only one $k$-IFFT at the transmitter and one $k$-FFT at the receiver is required. UFMC is thus more complex than OFDM, and this was to be expected since UFMC provides additional features with respect to OFDM. The increased complexity however can be easily managed with current state-of-the-art hardware technology.

\subsection{MIMO mmWave channel model}
As already discussed, at mmWave the MIMO channel model differs from the one usually employed at sub-6 GHz frequencies
\cite{buzzi2017massive, bjornson2018massive}. We provide here a brief description of the used channel model and 
refer the reader to   \cite{buzzidandreachannelmodel} for a complete specification of all its parameters.  
Denoting by $N_R$ and $N_T$ the number of receive and transmit antennas, respectively, 
the propagation channel can be modeled as an  $(N_R \times N_T)$-dimensional matrix-valued continuous time function, that we denote by $\mathbf{H}(t)$. 
According to the popular clustered model for MIMO mmWave channels, we assume that the propagation environment is made of $N_{\rm cl}$ scattering clusters, each of which contributes with $N_{{\rm ray}, i}$ propagation paths  $i=1, \ldots, {N_{\rm cl}}$, plus a 
possibly present LOS component. 
We denote by  $\phi_{i,l}^r$ and $\phi_{i,l}^t$ the azimuth angles of arrival and departure of the $l^{th}$ ray in the $i^{th}$ scattering cluster, respectively; similarly, $\theta_{i,l}^r$ and $\theta_{i,l}^t$ are the elevation angles  of arrival and departure of the $l^{th}$ ray in the $i^{th}$ scattering cluster, respectively. 
The impulse response  $\mathbf{H}(t)$ can be thus written as
\begin{multline}
{\mathbf{H}}(t)=\gamma\sum_{i=1}^{N_{\rm cl}}\sum_{l=1}^{N_{{\rm ray},i}}\alpha_{i,l}
\sqrt{L(r_{i,l})} \mathbf{a}_r(\phi_{i,l}^r,\theta_{i,l}^r) 
\cdot \\ 
\mathbf{a}_t^H(\phi_{i,l}^t,\theta_{i,l}^t)  \delta(t-\tau_{i,l}) + 
{\mathbf{H}}_{\rm LOS}(t)\; .
\label{eq:channel1}
\end{multline}
In the above equation,   $\alpha_{i,l}$ and $L(r_{i,l})$ are the complex path gain and the attenuation associated  to the $(i,l)$-th propagation path (whose length is denoted by $r_{i,l}$),  respectively; $\tau_{i,l}=r_{i,l}/c$, with $c$ the speed of light, is the propagation delay associated with the $(i,l)$-th path. 
The complex gain  $\alpha_{i,l}\thicksim \mathcal{CN}(0, \sigma_{\alpha,i}^2)$, with  $\sigma_{\alpha,i}^2=1$. 
The factors $\mathbf{a}_r(\phi_{i,l}^r,\theta_{i,l}^r)$ and $\mathbf{a}_t(\phi_{i,l}^t,\theta_{i,l}^t)$ represent the normalized receive and transmit array response vectors evaluated at the corresponding angles of arrival and departure; additionally, $\gamma=\displaystyle\sqrt{\frac{N_R N_T}{\sum_{i=1}^{N_{\rm cl}}N_{{\rm ray},i}}}$  is a normalization factor ensuring that the received signal power scales linearly with the product $N_R N_T$. 
Additionally, denoting by 
$\phi_{\rm LOS}^r$,  $\phi_{\rm LOS}^t$,
$\theta_{\rm LOS}^r$,  and $\theta_{\rm LOS}^t$ the departure angles corresponding to the LOS link, the LOS component is
\begin{equation}
\begin{array}{llll}
{\mathbf{H}}_{\rm LOS}(t) = &  
I_{\rm LOS}(d) \sqrt{N_R N_T} e^{j \eta} \sqrt{L(d)}\mathbf{a}_r(\phi_{\rm LOS}^r,\theta_{\rm LOS}^r) 
\cdot \\ & 
\mathbf{a}_t^H(\phi_{\rm LOS}^t,\theta_{\rm LOS}^t) \delta(t - \tau_{\rm LOS}) \; .
\end{array}
\label{eq:Hlos}
\end{equation}
In the above equation, $\eta \thicksim \mathcal{U}(0 ,2 \pi)$, while $I_{\rm LOS}(d) $ is an {indicator function/Bernoulli random variable, equal to 1} if a LOS link  between transmitter and receiver exists -- see \cite{buzzidandreachannelmodel} for the full details.

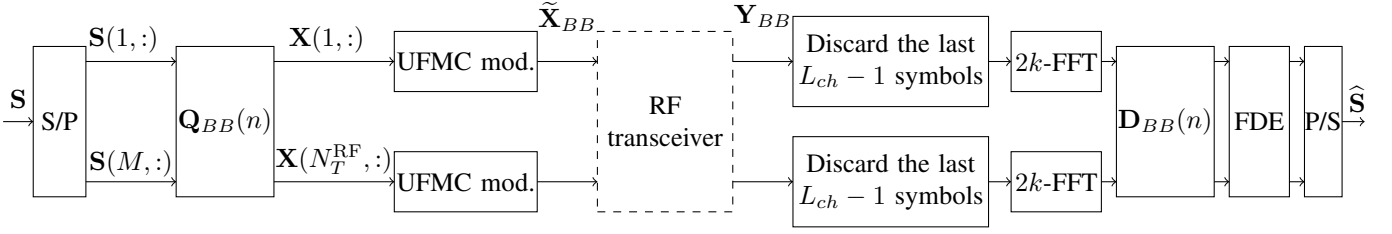
\begin{figure*}[h]
\begin{tikzpicture}[every text node part/.style={align=center}]
\draw [->](0,6-4) -- (0.4,6-4); \node at (0.2,6.3-4) {$\mathbf{S}$};
\draw (0.6-0.2,5-4)  rectangle (1.3-0.2,7-4) node[pos=0.5] {S/P} ;
\draw[->] (1.3-0.2,6.8-4) -- (2.5-0.2,6.8-4); \node at (1.8-0.2,7.1-4) {$\mathbf{S}(1,:)$};
\draw[->] (1.3-0.2,5.2-4) -- (2.5-0.2,5.2-4); \node at (1.9-0.2,5.45-4) {$\mathbf{S}(M,:)$};
\draw (2.5-0.2,5-4)  rectangle (3.8-0.2,7-4) node[pos=0.5] {$\mathbf{Q}_{BB}(n)$} ;
\draw[->] (3.8-0.2,6.8-4) -- (5.4-0.2,6.8-4); \node at (4.5-0.2,7.1-4) {$\mathbf{X}(1,:)$};
\draw[->] (3.8-0.2,5.2-4) -- (5.4-0.2,5.2-4); \node at (4.6-0.2,5.5-4) {$\mathbf{X}(N_T^{\rm RF},:)$};
\draw (5.4-0.2,6.4-4) rectangle (7.3-0.2,7.2-4) node[pos=0.5] {UFMC mod.};
\draw (5.4-0.2,4.8-4) rectangle (7.3-0.2,5.6-4) node[pos=0.5] {UFMC mod.};
\draw[->] (7.3-0.2,6.8-4) -- (8.2-0.3,6.8-4); \node at (7.8-0.3,7.4-4) {$\widetilde{\mathbf{X}}_{BB}$};
\draw[->] (7.3-0.2,5.2-4) -- (8.2-0.3,5.2-4); 
\draw [dashed] (8.2-0.3,4.8-4)  rectangle (10-0.3,7.2-4); \node at (9.1-0.3,6-4) {RF \\ transceiver};
\draw[->] (10-0.3,6.8-4) -- (10.8-0.3,6.8-4); \node at (10.4-0.3,7.4-4) {${\mathbf{Y}}_{BB}$};
\draw[->] (10-0.3,5.2-4) -- (10.8-0.3,5.2-4); 
\draw (13.5-0.4,6.2-4)  rectangle (10.8-0.3,7.4-4) node[pos=0.5, text width =2.7 cm] {Discard the last $L_{ch}-1$ symbols} ;
\draw (13.5-0.4,4.6-4)  rectangle (10.8-0.3,5.8-4) node[pos=0.5, text width =2.7 cm] {Discard the last $L_{ch}-1$ symbols} ;
\draw[->] (13.5-0.4,6.8-4) -- (13.8-0.4,6.8-4); 
\draw[->] (13.5-0.4,5.2-4) -- (13.8-0.4,5.2-4); 
\draw (13.8-0.4,6.4-4)  rectangle (15-0.4,7.2-4) node[pos=0.5] {$2k$-FFT} ;
\draw (13.8-0.4,4.8-4)  rectangle (15-0.4,5.6-4) node[pos=0.5] {$2k$-FFT} ;
\draw[->] (15-0.4,6.8-4) -- (15.2-0.4,6.8-4); 
\draw[->] (15-0.4,5.2-4) -- (15.2-0.4,5.2-4); 
\draw (15.2-0.4,5-4) rectangle (16.5-0.4,7-4) node [pos=0.5] {$\mathbf{D}_{BB}(n)$};
\draw[->] (16.5-0.4,6.8-4) -- (16.7-0.4,6.8-4); 
\draw[->] (16.5-0.4,5.2-4) -- (16.7-0.4,5.2-4); 
\draw (16.7-0.4,5-4) rectangle (17.5-0.4,7-4) node [pos=0.5] {FDE};
\draw[->] (17.5-0.4,6.8-4) -- (17.7-0.4,6.8-4); 
\draw[->] (17.5-0.4,5.2-4) -- (17.7-0.4,5.2-4); 
\draw (17.7-0.4,5-4) rectangle (18.2-0.4,7-4) node [pos=0.5] {P/S};
\draw[->] (18.2-0.4,6-4) -- (18.5-0.4,6-4); 
\node at (18.4-0.4,6.3-4) {$\widehat{\mathbf{S}}$};
\end{tikzpicture}
\caption{Block scheme of the UFMC multi antenna transceiver. The dashed box "RF transceiver" contains the cascade of a bank of $N_T^{\rm RF}$ transmit shaping filters, a bank of $N_T^{\rm RF}$ power amplifiers, the analog RF precoding matrix $\mathbf{Q}_{\rm RF}$, the $N_T$ transmit antennas, the $(N_R \times N_T)$-dimensional matrix-valued MIMO channel impulse response, the $N_R$ receive antennas, the analog RF postcoding matrix $\mathbf{D}_{\rm RF}$, and a bank of $N_R^{\rm RF}$
receive shaping filters.} 
\label{fig:UFMC_scheme_multiantenna}
\end{figure*}

\begin{table}[]
\caption{Meaning of the main mathematical symbols}
\label{Symbols_table}
\begin{tabular}{|l|l|}
\hline
\textbf{Symbol} & \textbf{Interpretation} \\ \hline
$\mathbf{S}$ & matrix contains informational symbols \\ \hline
$\hat{\mathbf{S}}$ & \begin{tabular}[c]{@{}l@{}}estimation of the matrix that contains \\ the informational symbols\end{tabular} \\ \hline
$\mathbf{W}_{k, IFFT}$, $\mathbf{W}_{k, FFT}$ & $k$-points IFFT and FFT matrices \\ \hline
$\mathbf{Q}_{BB}(n)$, $\mathbf{D}_{BB}(n)$ & \begin{tabular}[c]{@{}l@{}}digital precoding and postcoding \\ matrices on the $n$-th subcarrier\end{tabular} \\ \hline
$\mathbf{X}$ & \begin{tabular}[c]{@{}l@{}}matrix containing information \\ symbols after the digital precoding\end{tabular} \\ \hline
$\mathbf{P}_i$ & \begin{tabular}[c]{@{}l@{}} subcarriers selection matrix subcarrier \\ in the $i$-th subband\end{tabular} \\ \hline
$\mathbf{G}_i$ & \begin{tabular}[c]{@{}l@{}}toeplitz matrix describing the discrete\\ convolution operation\\ with the $i$-th prototype filter\end{tabular} \\ \hline
$\widetilde{\mathbf{X}}_{BB}$ & \begin{tabular}[c]{@{}l@{}}matrix at the output of \\ the MIMO-UFMC modulator\end{tabular} \\ \hline
$\mathbf{Q}_{\rm RF}$, $\mathbf{D}_{\rm RF}$ & digital precoding and postcoding matrices \\ \hline
$\widetilde{\mathbf{H}}(\ell)$ & \begin{tabular}[c]{@{}l@{}}matrix that contains the $\ell$-th sample \\ of the MIMO channel\end{tabular} \\ \hline
$\mathbf{Y}_{BB}$ & matrix at the output of the RF transceiver \\ \hline

\end{tabular}
\end{table}

\section{Transceiver schemes for MIMO-UFMC modulation} \label{UFMC_transceiver}

Equipped with the mathematical model of the single-antenna UFMC transceiver and with the above illustrated mmWave MIMO channel model, we are now ready to address the design of MIMO UFMC transceivers at mmWave. 
We will refer to the scheme reported in Fig. \ref{fig:UFMC_scheme_multiantenna}. In order to facilitate the reader, we report in Table \ref{Symbols_table} the meaning of the main mathematical symbols used in the paper.
We consider a MIMO single-user transmitter-receiver pair that, for an idealized scenario with a strictly orthogonal access scheme and no out-of-cell interference may be also representative of either the uplink or the downlink of a cellular system. The content of this paper can be generalized with ordinary efforts to the case a multiuser wireless system with co-channel interference; however, for the sake of simplicity this situation is not considered here.
We denote by $M$ the multiplexing order, and by $N_T^{\rm RF}<N_T$ and $N_R^{\rm RF}<N_R$ the number of RF chains at the transmitter and at the receiver, respectively. 
Focusing, for the moment, on single packet transmission, 
consider $Mk$ data symbols; these symbols are arranged into an $(M\times k)$-dimensional matrix, that we denote by $\mathbf{S}$. The columns of $\mathbf{S}$ undergo a digital precoding transformation; in particular, denoting by $\mathbf{Q}_{BB}(n)$ the $(N_T^{\rm RF} \times M )$-dimensional matrix representing the digital precoder (to be specified in the following) for the $n$-th column of $\mathbf{S}$, the useful data at the output of the digital precoding stage can be represented by the $(N_T^{\rm RF} \times k)$-dimensional matrix $\mathbf{X}$, whose $n$-th column, $\mathbf{X}(:,n)$ say, is expressed as
$\mathbf{X}(:,n)= \mathbf{Q}_{BB}(n) \mathbf{S}(:,n)$.
After digital precoding, each of the $N_T^{\rm RF}$ rows of the matrix $\mathbf{X}$ goes through an UFMC modulator, as the one depicted in Fig. \ref{fig:UFMC_scheme_singleantenna}; the outputs of the  $N_T^{\rm RF}$ parallel UFMC modulators  can be grouped in the matrix $\widetilde{\mathbf{X}}_{BB}$ of dimension $[N_T^{\rm RF} \times (k+L-1)]$. 
The $\ell$-th row of the output matrix $\widetilde{\mathbf{X}}_{BB}$ can be shown to be written as
\begin{equation}
\widetilde{\mathbf{X}}_{BB}(\ell,:)^T\!\!=\!\! \ds \sum_{i=0}^{B-1}\mathbf{G}_i \mathbf{W}_{k, IFFT} \mathbf{P}_i 
\mathbf{X}(\ell,:)^T, 
\label{eq:UFMCMIMO}
\end{equation} 
Eq. \eqref{eq:UFMCMIMO} can be compactly written in matrix notations as
$
\widetilde{\mathbf{X}}_{BB}=\mathbf{X} \left( \sum_{i=0}^{B-1}\mathbf{P}_i ^T \mathbf{W}_{k-IFFT}^T 
\mathbf{G}_i ^T \right).
$
The columns of $\widetilde{\mathbf{X}}_{BB}$ are then fed to the MIMO RF transceiver scheme,
that is made of the receive and transmit shaping filters, the analog precoding and postcoding matrices 
$\mathbf{Q}_{\rm RF}$ (of dimension  $N_T\times N_T^{\rm RF}$) and 
$\mathbf{D}_{\rm RF}$ (of dimension  $N_R\times N_R^{\rm RF}$), respectively, and of the MIMO channel impulse response. Assuming that the power amplifiers operate in the linear regime, the RF transceiver block can be modeled as an LTI filter with $(N_R^{\rm RF}
\times N_T^{\rm RF})$-dimensional matrix-valued impulse response $\mathbf{L}(\ell)=\sqrt{\frac{P_T}{M}}\mathbf{D}^H_{\rm RF} \widetilde{\mathbf{H}}(\ell) \mathbf{Q}_{\rm RF}$, wherein 
${P_T}$ is the transmitted power, 
$\widetilde{\mathbf{H}}(\ell)$, 
the sampled version of $\mathbf{H}(t)$, 
with $\ell=0,\ldots, L_{ch}-1$, is the matrix-valued $(N_R \times N_T)$-dimensional millimeter wave (mmWave) channel impulse response including also the transmit and receive rectangular shaping filters \cite{Buzzi_SCM2017}, with $L_{ch}$ the length of the channel impulse response (in discrete samples). The output of the RF transceiver can be represented through a matrix $\mathbf{Y}_{BB}$ of dimension $[N_R^{\rm RF} \times (k+L+L_{ch}-2)]$.  The $m$-th column of  $\mathbf{Y}_{BB}$ is easily seen to be expressed as
\begin{equation}
\mathbf{Y}_{\!BB}(:,m)\!\!=\! \mathbf{D}^H_{\rm RF}\!\! \left[\!\sum_{\ell=0}^{L_{ch}-1}\!\!\!\! \sqrt{\frac{P_T}{M}}  \widetilde{\mathbf{H}}(\ell) \mathbf{Q}_{\rm RF} \widetilde{\mathbf{X}}_{\!BB}(:,m\!-\!\ell)   \!+ \! \mathbf{w}(m) \!\right] \!,
\label{Y_BB}
\end{equation}
where we have assumed that $\widetilde{\mathbf{X}}(:,m)$ is zero for $m\leq 0$,
and the vector $\mathbf{w}(m)$ represents the additive thermal noise contribution.  Given the observable data  $\mathbf{Y}_{\!BB}$,  we now present some possible receiver algorithms. First of all, 
  following the usual UFMC processing, the last $L_{ch}-1$ columns of the matrix $\mathbf{Y}_{BB}$ are discarded, and each row of the resulting matrix, say $\widetilde{\mathbf{Y}}_{BB}$, is passed through an FFT on $2k$ points. The output of the FFT is downsampled by a factor of 2, so that we get a matrix of dimension 
$N_R^{\rm RF} \times k$, and finally, digital postcoding is applied. Denoting by $\mathbf{D}_{BB}(n)$ the $(N_R^{\rm RF}
\times M)$-dimensional matrix representing the digital postcoder (to be specified in the following) for the $n$-th column of the data matrix, we finally get a $(M \times k)$-dimensional matrix $\mathbf{Y}_{dec}$, whose $n$-th column can be shown to be approximately expressed as
\begin{multline}
\mathbf{Y}_{dec}(:,n) \approx \ds \frac{2k}{\sqrt{2}}\sqrt{\frac{P_T}{M}} \mathbf{D}_{BB}^H(n) \mathbf{D}^H_{\rm RF}
\bm{\overline{\mathcal{H}}}(2n-1) \mathbf{Q}_{RF} \times \\   
\widetilde{\mathbf{g}}_{\lfloor n/D \rfloor, (2n-1)}  
\mathbf{Q}_{BB}(n)\mathbf{S}(:,n) + 
\mathbf{D}_{BB}^H(n) \mathbf{D}^H_{\rm RF}  \bm{\mathcal{W}}(:,2n-1) \; ,
\label{eq:Y_dec}
\end{multline}
with $n=1, \ldots, k$. In the above equation, 
$\bm{\overline{\mathcal{H}}}(\ell)$ is an $N_R \times N_T$ matrix whose $(p,q)$-th entry is the $\ell$-th coefficient of the isometric $2k$-point FFT of the sequence  $\widetilde{\mathbf{H}}_{p,q}(0), \ldots,  
\widetilde{\mathbf{H}}_{p,q}(L_{ch}-1)$; similarly,   $\bm{\mathcal{W}}(:,\ell)$ is the $\ell$-th column of the matrix that contains the $2k$-point FFT of the  $\left[N_R \times (k+L+L_{ch}-2)\right]$-dimensional matrix defined as $\mathbf{W}_{\rm N}=\left[ \mathbf{w}(1) , \ldots , \mathbf{w}\left(k+L+L_{ch}-2\right) \right]$.
Now, given \eqref{eq:Y_dec}, an estimate of the $n$-th column of the data symbols matrix $\mathbf{S}$ can be simply obtained as
\begin{equation}
\widehat{\mathbf{S}}_{\rm id}(:,n) = \mathbf{E}(n) \mathbf{Y}_{dec}(:,n) \; ,
\label{eq:UFMC_decoding_MIMO}
\end{equation}
where 
$$
\begin{array}{lll}
\mathbf{E}(n)=& \left[\ds \frac{2k}{\sqrt{2}}\sqrt{\frac{P_T}{M}} \mathbf{D}_{BB}^H(n) \mathbf{D}^H_{\rm RF}
\bm{\overline{\mathcal{H}}}(2n-1)  \right. \\ & \left.  \mathbf{Q}_{RF}  
\widetilde{\mathbf{g}}_{\lfloor n/D \rfloor, (2n-1)}  
  \mathbf{Q}_{BB}(n)\right]^{+} \; .
\end{array}
$$
A different processing can be obtained by avoiding the use of the approximate relation \eqref{eq:Y_dec}. We thus consider the received matrix $\mathbf{Y}_{BB}$ in Eq. \eqref{Y_BB}, discard the last $L_{ch}-1$ columns of the matrix, and compute the FFT on $2k$ points: 
\begin{equation}
\mathbf{Y}_{{\rm dis}} = \mathbf{Y}_{BB} \mathbf{D}^T_{L_{ch}-1} \mathbf{W}_{2k, FFT}(1:k+L-1,:) \; .
\label{eq:Y_dec_real}
\end{equation}
An estimate of the $n$-th column of the data symbols matrix $\mathbf{S}$ can be thus obtained as
\begin{equation}
\widehat{\mathbf{S}}_{\rm dis}(:,n) = \mathbf{E}(n) \mathbf{Y}_{{\rm dis}}(:,2n-1) \;. 
\label{eq:UFMC_decoding_MIMO_real}
\end{equation}

We can also avoid discarding the last $L_{ch}-1$ columns of $\mathbf{Y}_{BB}$; in this case we obtain the processing
\begin{equation}
\widehat{\mathbf{S}}_{\rm no \; dis}(:,n) = \mathbf{E}(n)\mathbf{Y}_{BB} \mathbf{W}_{2k, FFT}(1:k+L+L_{ch}-2,2n-1) \; .
\label{eq:UFMC_decoding_MIMO_real_no_discard}
\end{equation}
Eqs. \eqref{eq:UFMC_decoding_MIMO_real} and \eqref{eq:UFMC_decoding_MIMO_real_no_discard}, have to be computed for  $n=1, \ldots, k$.

\subsection{Channel dependent beamforming (CDB)} \label{beamformers_CD}
We now address the beamformers choice by referring to  \eqref{eq:Y_dec}, which shows that
the precoding matrices multiply by the right the FFT channel coefficient $\bm{\overline{\mathcal{H}}}(2n-1)$, while the postcoding matrices multiply this same coefficient by the left. 
Denoting by $\mathbf{Q}^{\rm opt}(n)$ and $\mathbf{D}^{\rm opt}(n)$ the ``optimal'' precoding and postcoding matrices\footnote{By the adjective ``optimal'' we mean here the beamforming matrices that we would use in the case in which the number of RF chains coincides with the number of antennas.} for the transmission and detection of the $n$-th column of $\mathbf{S}$, it is seen from  \eqref{eq:Y_dec}  that, upon letting $\bm{\overline{\mathcal{H}}}(\ell) =\overline{\mathbf{U}}(\ell)\overline{\mathbf{\Lambda}}(\ell)\overline{\mathbf{V}}^H(\ell)$ be the singular-value-decomposition of $\bm{\overline{\mathcal{H}}}(\ell)$, the matrix $\mathbf{Q}^{\rm opt}(n)$ should contain on its columns the $M$ columns of  $\overline{\mathbf{V}}(2n-1)$
associated with the largest eigenvalues of $\bm{\overline{\mathcal{H}}}(2n-1) $, and, similarly, the matrix $\mathbf{D}^{\rm opt}(n)$ should contain on its columns the $M$ columns of $\overline{\mathbf{U}}(2n-1)$ associated with the largest eigenvalues of $\bm{\overline{\mathcal{H}}}(2n-1)$. 
Notice that the illustrated beamformers depend on the channel impulse response, whose knowledge is thus assumed at this stage: this explains why we use the acronym CDB to denote these beamformers. 

Now, given the matrices $\mathbf{Q}^{\rm opt}(n)$ and $\mathbf{D}^{\rm opt}(n)$, 
we need a procedure for finding the beamformers $\mathbf{Q}_{\rm BB}(n)$,
$\mathbf{Q}_{\rm RF}$, $\mathbf{D}_{ \rm BB}(n)$, and 
$\mathbf{D}_{\rm RF}$ reported in Fig. \ref{fig:UFMC_scheme_multiantenna}, so as to take into account the fact that the number of RF chains is smaller than the number of effective antennas, and, thus, hybrid analog/digital beamforming is to be used. 
Several algorithms are available in the open literature for approximating a given desired beamformer with an 
hybrid analog/digital structure; in this paper, we use the iterative 
approximation algorithm reported in \cite{ghauch2016}, omitting further details for the sake of brevity.

\section{MIMO-UFMC scheme with linear MMSE equalization at the receiver} \label{MMSE_UFMC_section}
\begin{figure}[h]
\centering
\includegraphics[scale=0.23]{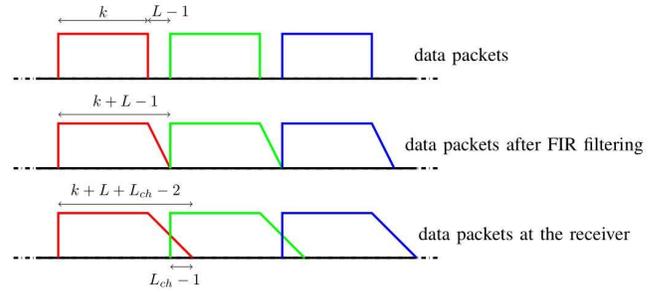}
\caption{Temporal separation (in discrete time) of the data packets in UFMC.}
\label{fig:UFMC_temporalpacketseparation}
\end{figure}

The transceiver processing described in the previous section requires the knowledge of the channel impulse response and is suited for a single packet transmission, i.e. for the case in which a single isolated block of $k$ symbols is transmitted. In practice, however, especially when considering massive broadband connections, several blocks are to be continuously transmitted. In this case, consecutive UFMC blocks are usually spaced in discrete-time by a number of intervals equal to $L-1$; this choice guarantees that the signals corresponding to contiguous blocks at the output of the Dolph-Chebyshev subband filtering do not interfere, so that no interblock interference (IBI) takes place at the transmitter side. However, since the channel is time-dispersive, at the receiver there will be IBI: in particular, the first $L_{ch}-1$ samples of the received vector $\widetilde{\mathbf{y}}$ in Eq. \eqref{eq:ytilde_UFMC} will be corrupted by the tail of the preceding block of data symbols -- see Fig \ref{fig:UFMC_temporalpacketseparation} for a graphical representation of the described situation. In this case, the single packet  processing described in the previous section is suboptimal and alternative interference-suppressing schemes are to be envisaged. 
In the following, 
we thus describe a linear MMSE-based processing operating directly on the matrix $\mathbf{Y}_{BB}$ reported in \eqref{Y_BB} and suited for multiple packet  transmission and reception, with and without the insertion of guard times among consecutive packets. 

\subsection{Channel independent  beamforming (CIB)} \label{beamformers_CI}
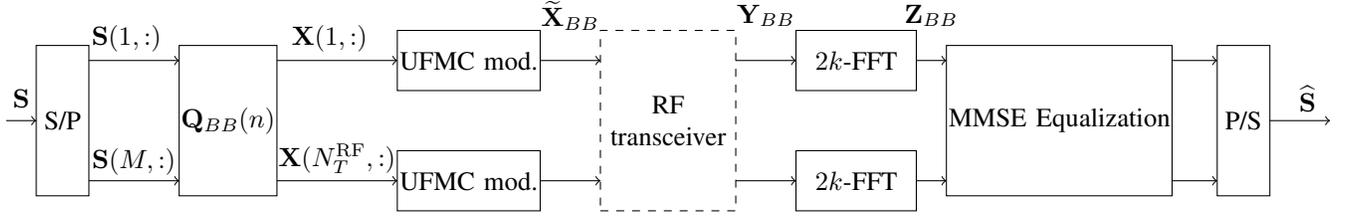
\begin{figure*}[h]
\begin{tikzpicture}[every text node part/.style={align=center}]
\draw [->](0,6-4) -- (0.4,6-4); \node at (0.2,6.3-4) {$\mathbf{S}$};
\draw (0.6-0.2,5-4)  rectangle (1.3-0.2,7-4) node[pos=0.5] {S/P} ;
\draw[->] (1.3-0.2,6.8-4) -- (2.5-0.2,6.8-4); \node at (1.8-0.2,7.1-4) {$\mathbf{S}(1,:)$};
\draw[->] (1.3-0.2,5.2-4) -- (2.5-0.2,5.2-4); \node at (1.9-0.2,5.45-4) {$\mathbf{S}(M,:)$};
\draw (2.5-0.2,5-4)  rectangle (3.8-0.2,7-4) node[pos=0.5] {$\mathbf{Q}_{BB}(n)$} ;
\draw[->] (3.8-0.2,6.8-4) -- (5.4-0.2,6.8-4); \node at (4.5-0.2,7.1-4) {$\mathbf{X}(1,:)$};
\draw[->] (3.8-0.2,5.2-4) -- (5.4-0.2,5.2-4); \node at (4.6-0.2,5.5-4) {$\mathbf{X}(N_T^{\rm RF},:)$};
\draw (5.4-0.2,6.4-4) rectangle (7.3-0.2,7.2-4) node[pos=0.5] {UFMC mod.};
\draw (5.4-0.2,4.8-4) rectangle (7.3-0.2,5.6-4) node[pos=0.5] {UFMC mod.};
\draw[->] (7.3-0.2,6.8-4) -- (8.2-0.3,6.8-4); \node at (7.8-0.3,7.4-4) {$\widetilde{\mathbf{X}}_{BB}$};
\draw[->] (7.3-0.2,5.2-4) -- (8.2-0.3,5.2-4); 
\draw [dashed] (8.2-0.3,4.8-4)  rectangle (10-0.3,7.2-4); \node at (9.1-0.3,6-4) {RF \\ transceiver};
\draw[->] (10-0.3,6.8-4) -- (10.8-0.3,6.8-4); \node at (10.4-0.3,7.4-4) {${\mathbf{Y}}_{BB}$};
\draw[->] (10-0.3,5.2-4) -- (10.8-0.3,5.2-4); 
\draw (12.5-0.4,6.4-4)  rectangle (10.8-0.3,7.2-4) node[pos=0.5] {$2k$-FFT} ;
\draw (12.5-0.4,4.8-4)  rectangle (10.8-0.3,5.6-4) node[pos=0.5] {$2k$-FFT} ;
 \node at (12.4-0.1,7.4-4) {${\mathbf{Z}}_{BB}$};
\draw[->] (12.5-0.4,6.8-4) -- (12.9-0.4,6.8-4); 
\draw[->] (12.5-0.4,5.2-4) -- (12.9-0.4,5.2-4); 
\draw (12.9-0.4,5-4) rectangle (15.9-0.4,7-4) node [pos=0.5] {MMSE Equalization};
\draw[->] (15.9-0.4,6.8-4) -- (16.5-0.4,6.8-4); 
\draw[->] (15.9-0.4,5.2-4) -- (16.5-0.4,5.2-4); 
\draw (16.5-0.4,5-4) rectangle (17.2-0.4,7-4) node [pos=0.5] {P/S};
\draw[->] (17.2-0.4,6-4) -- (18-0.4,6-4); 
\node at (17.7-0.4,6.3-4) {$\widehat{\mathbf{S}}$};
\end{tikzpicture}
\caption{Block scheme of the MIMO-UFMC MMSE-based multi antenna transceiver.} 
\label{fig:UFMC_scheme_multiantenna_MMSE_based}
\end{figure*}

First of all, we comment on the beamformers choice. As seen from the scheme of Fig. \ref{fig:UFMC_scheme_multiantenna_MMSE_based} the MMSE equalization block  incorporates the digital baseband beamformer at the receiver, while a channel-independent beamformer is used at the transmitter and in the RF section of the receiver so as to avoid the explicit need of channel state information. 
We thus  propose a  channel-independent beamforming scheme that can be easily implemented through the use of 0-1 switches. To this end, we assume, for the sake of simplicity, that the ratios
$N_T^{\rm RF}/M$,  $N_T/N_T^{\rm RF}$, and $N_R/N_R^{\rm RF}$ are integer numbers. 
In particular, the digital precoding $(N_T^{\rm RF} \times M)$-dimensional matrices are 
\begin{equation}
\mathbf{Q}_{\rm BB}^{\rm CI}(n)= \mathbf{I}_M \otimes \mathbf{1}_{N_T^{\rm RF}/M} \; \; \forall n=1, \ldots , k \; ,
\end{equation}
where $\mathbf{I}_M$ is the $(M \times M)$-dimensional identity matrix and $\mathbf{1}_{N_T^{\rm RF}/M}$ is the $\frac{N_T^{\rm RF}}{M}$-dimensional vector whose entries are all equal to 1, and $\otimes$ denotes the Kronecker product. 
Notice also that the above defined digital precoding matrices are no longer dependent on the subcarrier index. 
The analog precoding $(N_T \times N_T^{\rm RF})$- dimensional matrix is 
\begin{equation}
\mathbf{Q}_{\rm RF}^{\rm CI}= \mathbf{I}_{N_T^{\rm RF}} \otimes \mathbf{1}_{N_T/N_T^{\rm RF}} \; ,
\end{equation}
and the analog postcoding $(N_R \times N_R^{\rm RF})$- dimensional matrix is 
\begin{equation}
\mathbf{D}_{\rm RF}^{\rm CI}= \mathbf{I}_{N_R^{\rm RF}} \otimes \mathbf{1}_{N_R/N_R^{\rm RF}} \; .
\end{equation}

\subsection{LMMSE receiver processing}
The receiver processing is adaptive and so it, based on the transmission of training packets, automatically learns the interference-suppressing detection matrix; as a consequence, the detection strategy that we are going to illustrate is suited for multiple packet transmission, either with a guard-time between them, as recommended in \cite{Vakilian_UFMC2013} and depicted in Fig. \ref{fig:UFMC_temporalpacketseparation}, or with no guard-time at all. 

To begin with, we notice that, given the matrix $\mathbf{Y}_{BB}$, upon letting
$\mathbf{y}_{BB}={\rm vec}(\mathbf{Y}_{BB})$, the LMMSE estimate of the $k$-th column of the data matrix $\mathbf{S}$ is obtained as:
\begin{equation}
\widehat{\mathbf{S}}(:,k)=\mathbb{E}\left[\mathbf{S}(:,k) \mathbf{y}_{BB}^H \right]
\mathbb{E} \left[ \mathbf{y}_{BB}\mathbf{y}_{BB}^H\right]^{-1} \mathbf{y}_{BB} \; .
\label{eq:LMMSE_optimal_estimate}
\end{equation}
The above equation, however, can be hardly implemented in practice for two reasons. First, it requires accurate knowledge of the channel impulse response and of the transmitted powers in order to be able to compute the statistical expectations in \eqref{eq:LMMSE_optimal_estimate}; and, then, it involves numerical computations on vectors and matrices with size $N_R^{\rm RF} (k+L+L_{ch}-2)$, which can be a fairly large number. The former problem can be circumvented by resorting to adaptive signal processing schemes, while the latter can be addressed by processing vectors of reduced size, i.e. by properly windowing the received data vector based on the data symbol that we are interested in decoding. To this end, in order to compact the contribution of each data symbol to few entries of the received data, we have to consider an FFT processing of the rows of the data matrix $\mathbf{Y}_{BB}$; let us denote  by 
$\mathbf{Z}_{\rm BB}$ the $\left(N_R^{\rm RF} \times 2k \right)$-dimensional matrix contains the $2k$-points FFT of the matrix $\mathbf{Y}_{\rm BB}$ in Eq. \eqref{Y_BB}, i.e.,
\begin{equation}
\mathbf{Z}_{\rm BB} = \mathbf{Y}_{BB} \mathbf{W}_{2k, FFT}(1:k+L+L_{ch}-2,:) \; .
\label{Z_BB}
\end{equation}
We denote by $J$ the  window size, i.e. the number of columns of the matrix $\mathbf{Z}_{\rm BB}$ that we use to decode the symbols transmitted on the generic subcarrier; otherwise stated to limit system complexity, we use a window of data of dimension $J N_R^{\rm RF}$.
In order to estimate the symbols transmitted on the $n$-th subcarrier, we consider the $\left(N_R^{\rm RF} \times J\right)$-dimensional matrix $\mathbf{Z}_{ {\rm BB}}^{(n)}$, namely $\mathbf{Z}_{ {\rm BB}}^{(n)}$ is made of $J$ columns judiciously selected from the full matrix $\mathbf{Z}_{\rm BB}$, as fully explained in Algorithm \ref{Sel_subcarrier}. 
We denote by $\mathbf{z}_{{\rm BB}}^{(n)}$ the vector-stacked version of  $\mathbf{Z}_{ {\rm BB}}^{(n)}$, i.e. $\mathbf{z}_{{\rm BB}}^{(n)}=\text{vec}\left(\mathbf{Z}_{ {\rm BB}}^{(n)}\right)$, and we consider the linear processing
\begin{equation}
\widehat{\mathbf{S}}_{\rm mmse}(:,n)=\mathbf{D}_{\rm mmse}^H(n)\mathbf{z}_{ {\rm BB}}^{(n)} \; .
\label{UFMC_MMSE_estimated_reduced}
\end{equation}
In order to determine the detection matrix $\mathbf{D}_{\rm mmse}(n)$, we assume that $N_{\rm cov}$ training packets are transmitted; letting $\mathbf{z}_{ {\rm BB}, \ell}$ be the data vector coming from the $\ell$-th data packet, we build the following time-averaged covariance matrices:
\begin{equation}
\widetilde{\mathbf{R}}_z=\frac{1}{{N}_{\rm cov}}\sum_{\ell=1}^{{N}_{ \rm cov}} {\mathbf{z}_{ {\rm BB}, \ell}\mathbf{z}_{ {\rm BB}, \ell}^H} \; ,
\label{sample_cov1_reduced}
\end{equation}
\begin{equation}
\widetilde{\mathbf{R}}_{zs}^{(n)}= \frac{1}{{N}_{\rm cov}}\sum_{\ell=1}^{{N}_{ \rm cov}} {\mathbf{z}_{ {\rm BB}, \ell}\mathbf{S}_{\ell}(:,n)^H} \; .
\label{sample_cov2_reduced}
\end{equation}
Using Algorithm \ref{Sel_subcarrier}, we extract from $\widetilde{\mathbf{R}}_z$ and $\widetilde{\mathbf{R}}_{zs}^{(n)}$ the matrices ${\mathbf{R}}_{z,n}$ and ${\mathbf{R}}_{zs,n}$, and, finally, build the detection matrix 
\begin{equation}
\mathbf{D}_{\rm mmse}(n)={\mathbf{R}}_{z,n}^{-1} {\mathbf{R}}_{zs,n} \; .
\label{eq:detection_matrix}   
\end{equation}

\begin{algorithm}[!t]

\caption{Procedure for the selection of the quantities $\widetilde{\mathbf{R}}_{z,n}$, $\widetilde{\mathbf{R}}_{zs,n}$ and $\mathbf{Z}_{ {\rm BB},n}$, so as to determine the $J$ columns of $\mathbf{Z}_{BB}$ that contain the most significant contribution from the symbols in the $n$-th column of the data matrix $\mathbf{S}$. The notation
$\widetilde{\mathbf{R}}_{z}(a:b)$ denotes selection of a submatrix of $\widetilde{\mathbf{R}}_{z}$ containing the entries whose column and row coordinates  are in the range $(a:b)$. 
}

\begin{algorithmic}[1]

\label{Sel_subcarrier}

\IF {$n==1$ \OR $n==k$}

\IF {$n==1$}

\STATE {$I_{{\rm min},1}=1$, $I_{{\rm max},1}=J-2$, $I_{{\rm min},2}=2k-1$, $I_{{\rm max},2}=2k$}

\ELSIF {$n==k$}

\STATE {$I_{{\rm min},1}=1$, $I_{{\rm max},1}=2$, $I_{{\rm min},2}=2k-J+3$, $I_{{\rm max},2}=2k$}

\ENDIF 

\STATE {$\mathbf{Z}_{ {\rm BB}}^{(n)}=\left[\mathbf{Z}_{\rm BB}\left(:,I_{{\rm min},1}:I_{{\rm max},1}\right) ,  \mathbf{Z}_{\rm BB}\left(:,I_{{\rm min},2}:I_{{\rm max},2}\right) \right]$.}

\STATE{${\mathbf{R}}_{z,n}=\left[\widetilde{\mathbf{R}}_{z}\left(N_R^{\rm RF}\left(I_{{\rm min},1}-1\right)+1:N_R^{\rm RF}I_{{\rm max},1}\right), \right. $} 

\STATE{$ \left. \widetilde{\mathbf{R}}_{z}\left(N_R^{\rm RF}\left(I_{{\rm min},2}-1\right)+1:N_R^{\rm RF}I_{{\rm max},2}\right) \right]$. }

\STATE{${\mathbf{R}}_{zs,n}=\left[\widetilde{\mathbf{R}}_{zs}^{(n)}\left(N_R^{\rm RF}\left(I_{{\rm min},1}-1\right)+1:N_R^{\rm RF}I_{{\rm max},1},:\right), \right.$ }

\STATE{$ \left. \widetilde{\mathbf{R}}_{zs}^{(n)}\left(N_R^{\rm RF}\left(I_{{\rm min},2}-1\right)+1:N_R^{\rm RF}I_{{\rm max},2},:\right) \right]$. }

\ELSE

	\IF { $ 2n-\frac{J}{2} \geq 1$ \AND $2n+\frac{J}{2}-1\leq 2k$}

 		\STATE {$I_{\rm min}=2n-\frac{J}{2}$, $I_{\rm max}=2n+\frac{J}{2}-1$}
	\ELSIF  { $ 2n-\frac{J}{2} < 1$}
		\STATE {$I_{\rm min}=1$, $I_{\rm max}=J$}
	\ELSIF  { $ 2n+\frac{J}{2}-1 > 2k$}
		\STATE {$I_{\rm min}=2k-J+1$, $I_{\rm max}=2k$}

\ENDIF

\STATE {$\mathbf{Z}_{ {\rm BB}}^{(n)}=\mathbf{Z}_{\rm BB}\left(:,N_R^{\rm RF}\left(I_{{\rm min}}-1\right)+1:N_R^{\rm RF} I_{\rm max}\right)$.}

\STATE{${\mathbf{R}}_{z,n}=\widetilde{\mathbf{R}}_{z}\left(N_R^{\rm RF}\left(I_{{\rm min}}-1\right)+1:N_R^{\rm RF}I_{{\rm max}}\right)$. }

\STATE{${\mathbf{R}}_{zs,n}=\widetilde{\mathbf{R}}_{zs}^{(n)}\left(N_R^{\rm RF}\left(I_{{\rm min}}-1\right)+1:N_R^{\rm RF}I_{{\rm max}},:\right)$. }

\ENDIF
\end{algorithmic}

\end{algorithm}


\subsection{LMMSE receiver processing with learning capabilities}
The detection matrix in \eqref{eq:detection_matrix} is based on time-averaged estimates of suitable covariance matrices, as shown in \eqref{sample_cov1_reduced} and \eqref{sample_cov2_reduced}; the considered time-averages equally weight the $N_{\rm cov}$ training packets, and, moreover, \eqref{eq:detection_matrix} entails a matrix inversion, a task that is usually computationally intensive. 
In time-varying environments, and in situations where receiver complexity is an issue, it is convenient to resort to adaptive and \textit{recursive} implementations of the LMMSE receiver. Indeed, recursive algorithms serve at least two purposes. First, they have a learning and tracking capability that weighs more the recent past, so that changes in the channel impulse response or in the interference can be fastly incorporated in the receiver; this thus results in the capability to learn the interference environment and, it its changes are relatively slow, to continuously adapt to it.
Second, they do not rely on direct matrix inversion as in \eqref{eq:detection_matrix} and so have a lower computational complexity.
In the considered scenario, each packet contains $MK$ data symbols, so $MK$ adaptive learning algorithms are to be run in parallel with iterations occurring on a timescale equal to the inverse of the packet frequency; we denote by $\mathbf{S}_{\ell}$ the data matrix for the $\ell$-th packet.
The procedures work as follows. Consider the vector $\mathbf{z}_{{\rm BB}, \ell}$ containing the $2k$-dimensional data from the $\ell$-th data packet. Assuming that we are interested to detect the data sequence that happens to be located on the $n$-th column of the data matrices $\ldots, \mathbf{S}_{\ell}, \mathbf{S}_{\ell+1}, \ldots$, use the procedure in Algorithm \ref{Sel_subcarrier} to obtain the useful $JN^{\rm RF}_R$-dimensional data vector, $\mathbf{z}_{{\rm BB},\ell}(n)$ say. The vectors 
$\ldots, \mathbf{z}_{{\rm BB},\ell}^{(n)}, \mathbf{z}_{{\rm BB},\ell+1}^{(n)}, \ldots$ can be thus used as an input to an adaptive algorithm in order to detect the data on the $n$-th column of the data matrices $\ldots, \mathbf{S}_{\ell}, \mathbf{S}_{\ell+1}, \ldots$. 
Several adaptive algorithms can be used to approximate the linear MMSE receiver. In the following we briefly report some details on the Normalized Least-Mean-Squares (NLMS)  and the Recursive-Least-Squares (RLS) algorithms
\cite{sayed2011adaptive}. For ease of notation, we denote by $d(\ell)$ the desired quantity in the $\ell$-th temporal interval (e.g., the $(m,k)$-th entry of $\mathbf{S}_{\ell}$), by $\mathbf{u}_{\ell}$ the $JN^{\rm RF}_R$-dimensional observable vector, in row-format,  to be processed in the $\ell$-th temporal interval, and by $\mathbf{w}_{\ell}$ the estimate of the optimal filter at the $\ell$-th temporal interval, so that the soft estimate of  $d(\ell)$ is obtained as $\mathbf{u}_{\ell} \mathbf{w}_{\ell}$. The NLMS  and the RLS procedures  are reported in Algorithm \ref{NLMS_Algorithm} and in Algorithm \ref{RLS_Algorithm}, respectively. 

\begin{algorithm}[!t]

\caption{NLMS Algorithm. Let $\mu$ be a positive step-size, $\epsilon$, a small positive parameter, $L_T$ the length of the training interval and denote by $\widehat{(\cdot)}$  the minimum-distance projection of $(\cdot)$ on the used modulation constellation.}

\begin{algorithmic}[1]

\label{NLMS_Algorithm}

\IF {$i \leq L_T$}

\STATE {$\mathbf{w}_i=\mathbf{w}_{i-1}+\frac{\mu}{\epsilon+||\mathbf{u}_i||^2}\mathbf{u}_i^H \left[d(i)-\mathbf{u}_i\mathbf{w}_{i-1}\right]$}

\ELSE 

\STATE {$\mathbf{w}_i=\mathbf{w}_{i-1}+\frac{\mu}{\epsilon+||\mathbf{u}_i||^2}\mathbf{u}_i^H \left[\widehat{\mathbf{u}_i\mathbf{w}_{i-1}}-\mathbf{u}_i\mathbf{w}_{i-1}\right]$}

\ENDIF 
\end{algorithmic}

\end{algorithm}

\begin{algorithm}[!t]

\caption{RLS Algorithm.  Let $\epsilon$ be a small positive parameter and $\lambda$ is a close to unity, but smaller than unity, constant, usually named the forgetting factor.}

\begin{algorithmic}[1]

\label{RLS_Algorithm}

\IF {$i \leq L_T$}

\STATE {$\mathbf{w}_i=\mathbf{w}_{i-1}+\frac{\lambda^{-1}\mathbf{P}_i \mathbf{u}_i^H}{1+\lambda^{-1}\mathbf{u}_i\mathbf{P}_{i-1}\mathbf{u}_i^H}  \left[d(i)-\mathbf{u}_i\mathbf{w}_{i-1}\right]$}

\ELSE 

\STATE {$\mathbf{w}_i=\mathbf{w}_{i-1}+\frac{\lambda^{-1}\mathbf{P}_i \mathbf{u}_i^H}{1+\lambda^{-1}\mathbf{u}_i\mathbf{P}_{i-1}\mathbf{u}_i^H} \left[\widehat{\mathbf{u}_i\mathbf{w}_{i-1}}-\mathbf{u}_i\mathbf{w}_{i-1}\right]$}

\ENDIF 

\STATE {$\mathbf{P}_i=\lambda^{-1}\left[\mathbf{P}_{i-1}-\frac{\lambda^{-1}\mathbf{P}_{i-1}\mathbf{u}_i^H\mathbf{u}_i\mathbf{P}_{i-1}}{1+\lambda^{-1}\mathbf{u}_i\mathbf{P}_{i-1}\mathbf{u}_i^H}\right]$}
\end{algorithmic}

\end{algorithm}

\noindent
\textit{Computational complexity.}
We now provide remarks on the complexity of the proposed procedures. Algorithm \ref{Sel_subcarrier} is simply an entry selection procedure that does not involve any computation; for each subcarrier, it simply selects the data that will be used in the following data decoding algorithms, so its complexity is negligible. To decode the data symbols in each packet, $MK$ learning algorithms are to be run in parallel. With regard to the NLMS algorithms, detailed in Algorithm \ref{NLMS_Algorithm}, the complexity is linear with the dimension of the input data\cite{sayed2011adaptive}, so that the complexity can be approximated as $O\left( MK J N_R^{\rm RF}\right)$ per decoded packet. For the case in which we use $MK$ RLS algorithms (detailed in Algorithm \ref{RLS_Algorithm}), the complexity of each algorithm is quadratic with the dimension of the input data\cite{sayed2011adaptive}, so that the complexity  can be approximated as $O\left[ MK \left( J N_R^{\rm RF}\right)^2 \right]$ per decoded packet.

\section{Performance measures, analysis, and numerical results}
We now provide numerical results showing the performance of the proposed MIMO-UFMC transceiver architectures. We will consider three different performance measures. The first one is the root mean square error (RMSE) defined as

\begin{equation}
\rm{RMSE}= \mathbb{E}\left[\frac{|s - \hat{s}|^2}{|s|^2}\right] \; ,
\label{RMSE}
\end{equation}
where $s$ and $\hat{s}$ are the generic symbol transmitted and estimated, respectively.

The second one is the usual BER, while, finally, the third one is the throughput, that 
 is measured in bit/s, and depends  on the system BER and on the cardinality of the used modulation. Denoting by $T_s$ the signaling time, i.e. assuming that the modulator transmits a data-symbol of cardinality $\cal M$ every $T_s$ seconds,  $kM$ symbols are transmitted in $(k+L-1)T_s$ seconds, if a guard interval is inserted among consecutive packets,  and in $k T_s$ seconds, in the case in which packets are continuously transmitted with no time-spacing among them. Denoting by $W$ the communication bandwidth, 
and assuming $T_s=1/W$,  
 the throughput ${\cal T}_{\rm G}$ for the former scenario can be written as 
\begin{equation}
{\cal T}_{\rm G} = \ds \frac{W \log_2\left( {\cal M} \right) k M}{k +L -1}(1-{\rm BER}) \; \, \quad [{\rm bit/s}] \; , 
\label{eq:throughput_G}
\end{equation}
while, instead, in the latter situation we have 
\begin{equation}
{\cal T}_{\rm NG} = \ds W \log_2\left( {\cal M} \right) M (1-{\rm BER}) \; \, \quad [{\rm bit/s}] \; . 
\label{eq:throughput_NG}
\end{equation}

\subsection{Mathematical Analysis} \label{GA_section}
We now provide a closed-form expression for the BER of the two main architectures detailed in the paper, i.e. the MIMO-UFMC transceiver of Section \ref{UFMC_transceiver} and the MIMO-UFMC scheme with MMSE equalization at the receiver of Section \ref{MMSE_UFMC_section}. For the closed form of the BER, we resort to the Gaussian Approximation (GA) of the interference \cite{poor1997probability}. Assuming the use of Gray coding, and  a
 $Q$-ary quadrature amplitude modulation (QAM), the BER is expressed as
 
\begin{equation}
\begin{array}{ll}
\text{BER}_{  Q-\rm{QAM}} =& \ds \frac{1}{\log_2 (Q)} \times \\ &
 \left( 1 - \left[ 1- \frac{\sqrt{Q}-1}{\sqrt{Q}} \text{erfc}\left( \sqrt{\frac{3 \, \text{SINR}}{2 (Q-1)}}\right)\right]^2\right),
\end{array}\label{BER_Q_QAM}
\end{equation}
 where SINR is the Signal-to-Interference-plus-Noise Ratio, to be computed in the following.

\medskip

Consider now the case of MIMO-UFMC processing.
Eq. \eqref{eq:UFMC_decoding_MIMO} can be written as 
\begin{equation}
\widehat{\mathbf{S}}_{\rm id}(:,n) = \mathbf{S}(:,n) + \mathbf{E}(n) \mathbf{D}_{BB}^H(n) \mathbf{D}^H_{\rm RF}  \bm{\mathcal{W}}(:,2n-1) \; ,
\label{eq:UFMC_decoding_MIMO2}
\end{equation}
with the noise contribution $\bm{\mathcal{W}}(:,2n-1)$ expressed as\cite{Buzzi_SCM2017}
\begin{equation}
\begin{array}{lll}
\bm{\mathcal{W}}(:,2n-1)= \\ \left[\mathbf{I}_{N_{\rm R}} \otimes \mathbf{W}_{2k, FFT}(2n-1,1:k+L+L_{ch}-2)\right]  \text{vec}\left( \mathbf{W}_{\rm N}^T\right).
\end{array}
\end{equation}
Denoting by $\mathbf{C}_{\rm W}$ the correlation matrix of the rows of the matrix $\mathbf{W}_{\rm N}$ that contains the noise, the SINR for the MIMO-UFMC processing on the $n$-th subcarrier, can be written as
\begin{equation}
\text{SINR}_{\rm MIMO-UFMC}^{(n)}= \frac{1}{\text{tr}\left( \mathbf{B}(n) \widetilde{\mathbf{C}}_{\rm W} \mathbf{B}(n)^H\right)} \; ,
\label{SNR_UFMC_GA}
\end{equation}
where $\widetilde{\mathbf{C}}_{\rm W}=\left[\mathbf{I}_{N_{\rm R}} \otimes \mathbf{C}_{\rm W}\right]$ and
\begin{equation}
\begin{array}{llll}
\mathbf{B}(n)=&\mathbf{E}(n) \mathbf{D}_{BB}^H(n) \mathbf{D}^H_{\rm RF}\times \\ & \left[\mathbf{I}_{N_{\rm R}} \otimes \mathbf{W}_{2k, FFT}(2n-1,1:k+L+L_{ch}-2)\right]\, .
\end{array}
\end{equation}

\

\medskip

Consider now the MIMO-UFMC scheme with linear MMSE equalization at the receiver, and in particular
 Eq. \eqref{UFMC_MMSE_estimated_reduced}, where we assume, for ease of notation, that $\mathbf{z}_{ {\rm BB}}^{(n)}=\mathbf{z}_{ {\rm BB}}=\text{vec}\left(\mathbf{Z}_{\rm BB}\right)$, so that we are considering all the columns of the matrix $\mathbf{Z}_{\rm BB}$ in order to decode the $M$ symbols in the $n$-th column of the data matrix $\mathbf{S}$. After some algebraic manipulations, using the properties of the vec$(\cdot)$ operator and of the Kronecker product, it can be shown that $\mathbf{z}_{ {\rm BB}}$ can be written as follows
\begin{equation}
\mathbf{z}_{ {\rm BB}}=\sqrt{\frac{P_T}{M}}\bar{\mathbf{A}}\mathbf{s}+\overline{\mathbf{B}}\overline{\mathbf{w}} \, ,
\label{z_BB_matrix_form}
\end{equation}
where $\mathbf{s}=\text{vec}\left( \mathbf{S}\right)$, $\overline{\mathbf{w}}=\text{vec}\left( \mathbf{W}_{\rm N}\right)$,
\begin{equation}
\overline{\mathbf{B}}= \mathbf{W}_{2k, FFT}(1:k+L+L_{ch}-2,:)^T \otimes \mathbf{D}_{\rm RF}^H \, ,
\label{B_bar}
\end{equation}
\begin{equation}
\begin{array}{llll}
\bar{\mathbf{A}}=&\left[ \mathbf{W}_{2k, FFT}(1:k+L+L_{ch}-2,:)^T \otimes \mathbf{I}_{N_{\rm R}^{\rm RF}} \right]\times \\&   \mathbf{A}_{ch}  \left( \mathbf{U}^T \otimes \mathbf{I}_{N_{\rm R}^{\rm RF}} \right) \widetilde{\mathbf{Q}}_{\rm BB} \, ,
\end{array}
\end{equation}
with $\mathbf{A}_{ch}$ is the convolution matrix defined starting from Eq. \eqref{Y_BB}, $\mathbf{U}=\left( \sum_{i=0}^{B-1}\mathbf{P}_i ^T \mathbf{W}_{k-IFFT}^T 
\mathbf{G}_i ^T \right)$, and $\widetilde{\mathbf{Q}}_{\rm BB}=\text{blkdiag}\left(\mathbf{Q}_{\rm BB}(1), \ldots, \mathbf{Q}_{\rm BB}(n) \right)$. 

Exploiting \eqref{z_BB_matrix_form}, \eqref{UFMC_MMSE_estimated_reduced} 
can be rewritten as
\begin{equation}
\begin{array}{lll}
\widehat{\mathbf{S}}_{\rm mmse}(:,n)=& \ds \sqrt{\frac{P_T}{M}}\mathbf{D}_{\rm mmse}^H(n)\bar{\mathbf{A}}_n \mathbf{S}(:,n) \\ &+ \ds \sqrt{\frac{P_T}{M}}\mathbf{D}_{\rm mmse}^H(n)\bar{\mathbf{A}}_{\bar{n}} \mathbf{s}_{\bar{n}} +\mathbf{D}_{\rm mmse}^H(n) \overline{\mathbf{B}}\overline{\mathbf{w}} \, ,
\end{array}
\label{UFMC_MMSE_estimated_reduced_matrix_form}
\end{equation}
where $\bar{\mathbf{A}}_n=\bar{\mathbf{A}}(:, (n-1)M+1: (n-1)M+M)$, $\bar{\mathbf{A}}_{\bar{n}}$ contains all the columns of the matrix $\bar{\mathbf{A}}$ except the ones in $\bar{\mathbf{A}}_n$ and $\mathbf{s}_{\bar{n}}$ contains all the entries of the vector $\mathbf{s}$ except the entries from the $[(n-1)M+1]$-th to the $[(n-1)M+M]$-th.
Based on \eqref{UFMC_MMSE_estimated_reduced_matrix_form}, the $n$-th subcarrier SINR for the case of MIMO-UFMC architecture with linear MMSE at the receiver can be finally written as
\begin{equation}
\begin{array}{lll}
\text{SINR}_{\rm UFMC-MMSE}^{(n)}= \\ \ds \frac{\frac{P_T}{M} \text{tr} \left( \mathbf{D}_{\rm mmse}^H(n)\bar{\mathbf{A}}_n\bar{\mathbf{A}}_n^H \mathbf{D}_{\rm mmse}(n)\right)}{ \frac{P_T}{M} \text{tr} \left( \mathbf{D}_{\rm mmse}^H(n)
\left( \bar{\mathbf{A}}_{\bar{n}}\bar{\mathbf{A}}_{\bar{n}}^H   + 
\overline{\mathbf{B}} \; \overline{\mathbf{C}}_{\rm W} \overline{\mathbf{B}}^H
\right)\mathbf{D}_{\rm mmse}(n)\right)} \; ,
\end{array}
\label{SNR_UFMC_MMSE_GA}
\end{equation}
with  $\overline{\mathbf{C}}_{\rm W}=\left[\mathbf{C}_{\rm W} \otimes \mathbf{I}_{N_{\rm R}}  \right]$.
Inserting \eqref{SNR_UFMC_GA} and \eqref{SNR_UFMC_MMSE_GA} into in \eqref{BER_Q_QAM} it is possible to have an approximate expression for the system BER, for the cases of MIMO-UFMC processing and of MIMO-UFMC with linear MMSE equalization at the receiver, respectively. 

\subsection{Numerical Results}
In order to produce numerical results, we consider a communication bandwidth of $W=500$ MHz centered over a mmWave carrier frequency. The MIMO propagation channel has been generated according to the statistical procedure described in\cite{Buzzi_SCM2017,buzzidandreachannelmodel}. We assume a distance between transmitter and receiver of 50 meters. The additive thermal noise is assumed to have a power spectral density of -174 dBm/Hz, while the front-end receiver is assumed to have a noise figure of 3 dB. For the prototype filter in the UFMC modulators we use a Dolph-Chebyshev filter with length $L=16$ and side-lobe attenuation with respect to the peak of the main lobe equal to 100 dB. We use $k=128$ subcarriers, and $B=8$ subbands (which leads to $D=16$ subcarriers in each subband). We consider the antenna configuration $N_R \times N_T = 16 \times 64$, and we assume hybrid beamforming with $N_T^{\rm RF}=16$ and $N_R^{\rm RF}=4$ RF chains.  For single packet transmission and reception, in the figures we denote as ``UFMC-id'' the case in which the estimate of the $n$-th column of the data symbols matrix $\mathbf{S}$ is expressed as \eqref{eq:UFMC_decoding_MIMO}, as ``UFMC-dis'' the case in which we use \eqref{eq:UFMC_decoding_MIMO_real}, as ``UFMC-no dis'' the case in which we use \eqref{eq:UFMC_decoding_MIMO_real_no_discard}, as ``UFMC-mmse'' the case in which we use \eqref{UFMC_MMSE_estimated_reduced}. With regard to the multiple packet transmission, we label with ``UFMC-mmse-G'' the case in which we adopt the LMMSE receive and consecutive packets are spaced apart by $L-1$ time intervals; finally we label with ``UFMC-mmse-NG'' the case in which the LMMSE receiver is adopted and packets are continuously transmitted. The acronyms SP and MP will be used to distinguish the case of isolated single-packet transmission from the case of multiple packet transmission. 
We will also consider the effect on the system performance of phase noise  generated by the local oscillators at the receiver. Details on the phase noise model are reported in the Appendix, while the {\tt PhaseNoise} System object™ in the Communications System Toolbox™ of MATLAB has been used to generate the phase noise in the numerical simulations. 

\begin{figure*}[t]
\begin{center}
\includegraphics[scale=0.57]{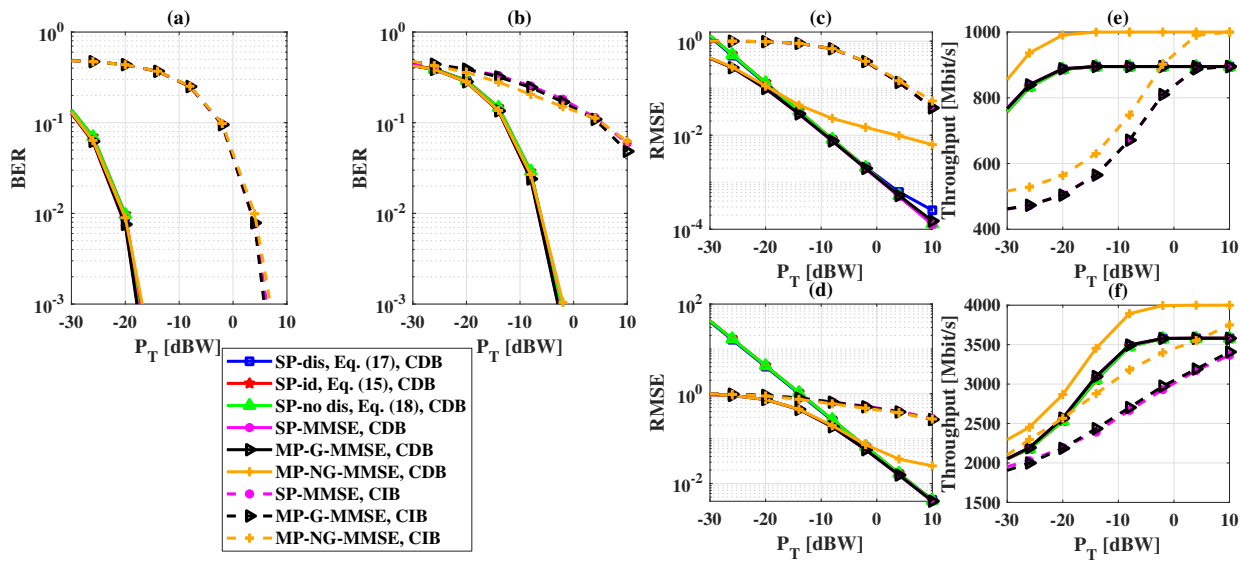}
\end{center}
\caption{Performance measures of  MIMO-UFMC transceiver architectures versus transmit power, with CDB and CIB,  4-QAM modulation, and no phase noise. Subfigure $(a)$: BER versus transmit power, $M=1$; subfigure $(b)$: BER versus transmit power, $M=4$; subfigure $(c)$: RMSE versus transmit power, $M=1$; subfigure $(d)$: RMSE versus transmit power, $M=4$; subfigure $(e)$: throughput versus transmit power, $M=1$; subfigure $(f)$: throughput versus transmit power, $M=4$.}
\label{Fig:MIMO_Performances_M14}
\end{figure*}

\begin{figure*}[t]
\begin{center}
\includegraphics[scale=0.5]{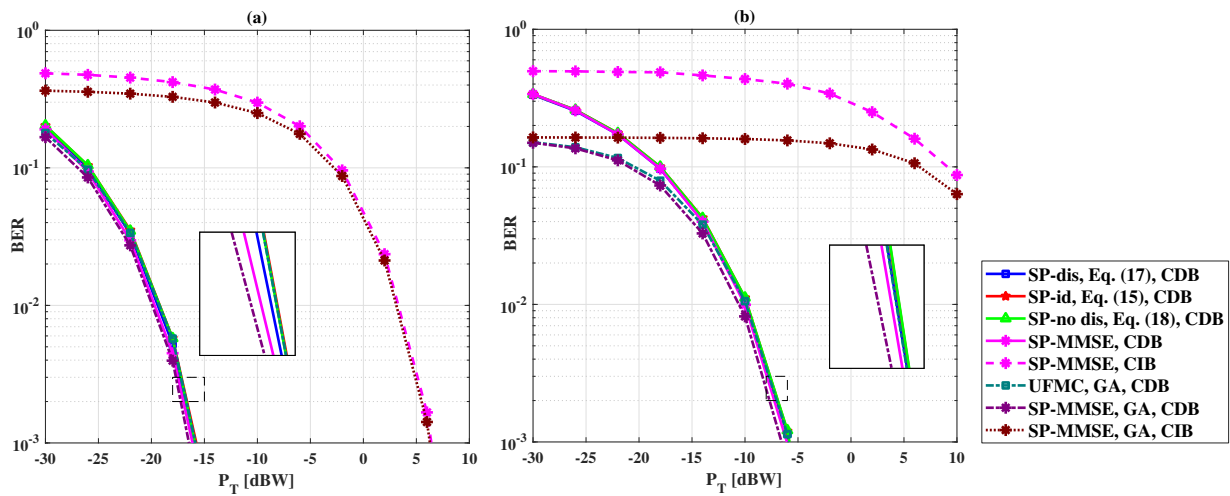}
\end{center}
\caption{Comparison between the BER performance obtained with simulations and with the GA of MIMO-UFMC transceiver architectures with CDB and CIB, no phase noise and single packet transmission. Subfigure $(a)$: 4-QAM modulation; subfigure $(b)$: 64-QAM modulation.}
\label{Fig:BER_GA_comparison}
\end{figure*}

\subsubsection{System performance with no phase noise}
Fig. \ref{Fig:MIMO_Performances_M14} shows the performance of the considered transceiver schemes neglecting the phase noise at the receiver. In particular, the figure reports the three discussed performance measures versus the transmit power, assuming 4-QAM modulation, and with two different values of the multiplexing order, that is $M=1$ and $M=4$. Both the cases of CDB and CIB are considered, as well as the cases of SP and MP transmission, with and without the guard time between consecutive packets.  The following comments can be done. First of all, from subplots (a) - (d) it is seen that there is a considerable performance gap between the case of CDB and of CIB: this behavior appears reasonable and can be justified by noticing that when using CIB no information on the channel coefficients is needed at the transmitter. Moreover, it is seen that when considering multiple packet transmission, including a guard time among consecutive packets has a very negligible effect on the system BER and RMSE; otherwise stated, the proposed MMSE receiver is capable of managing the increased inter-packet interference that arises when data packets are continuously transmitted with no spacing among them, achieving a BER and a RMSE almost indistinguishable from the case in which packets are spaced by a guard time. 
This, in turn, has a large impact on the system throughput, as shown in subplots (e) - (f). Indeed, it is seen that the removal of the guard time allows increasing the system throughput of about 13 - 15 $\%$, thus confirming that tolerating an increased interference level is compensated by a considerable increase in the system throughput. Subplots (e) - (f) also show that the structures with CIB eventually achieve, for large values of the transmitted power, the same throughput as the structures with CDB. 

Fig. \ref{Fig:BER_GA_comparison} is devoted to the validation of the BER 
approximation \eqref{BER_Q_QAM}. Indeed, the figure shows the system BER  in the case of single packet transmission for the considered MIMO-UFMC architectures, for both QPSK and 64-QAM modulation. It is seen that the GA provides a lower bound of the BER, expecially in the low-SINR regime, while in the region of interest of high-SNR regime it gives a very good approximation of the simulated BER. The observed gap for the low-SINR region is due to the fact that the $1/\log_2(Q)$ factor, coming from the Gray coding approximation, is too optimistic. In any case, what really matters is the fact that in the large-SINR region the GA works well, since in this region performing MonteCarlo simulations requires increasingly large large CPU times.

\subsubsection{Impact of phase noise}
\begin{figure*}[t]
\begin{center}
\includegraphics[scale=0.52]{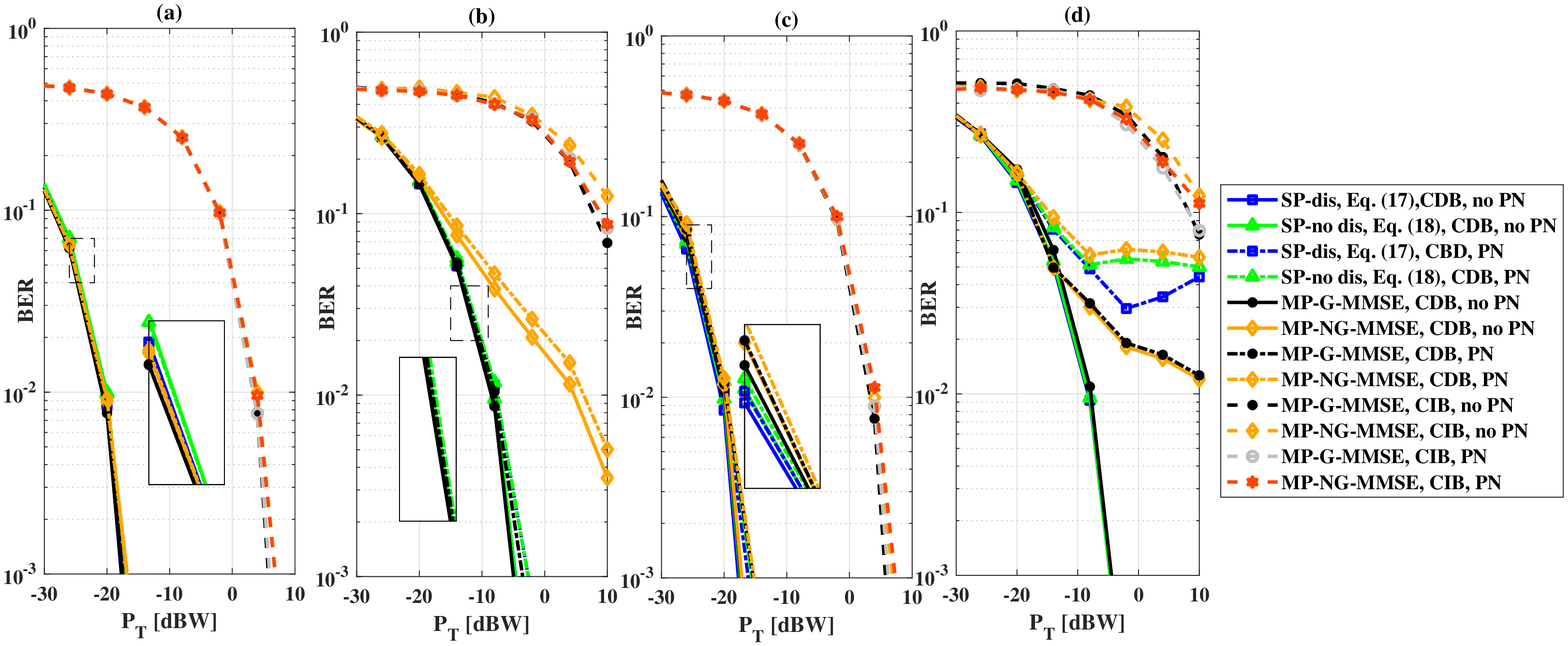}
\end{center}
\caption{BER versus transmit power, performance of MIMO-UFMC transceiver architectures with CDB and CIB, and phase noise at the receiver: in subfigure $(a)$ $L_{\rm PN}=-60$ dBc/Hz and $f_{\rm offset}=100$ Hz, 4-QAM modulation, in subfigure $(b)$ $L_{\rm PN}=-60$ dBc/Hz and $f_{\rm offset}=100$ Hz, 64-QAM modulation, in subfigure $(c)$ $L_{\rm PN}=-89$ dBc/Hz and $f_{\rm offset}=1$ MHz, 4-QAM modulation, in subfigure $(d)$ $L_{\rm PN}=-89$ dBc/Hz and $f_{\rm offset}=1$ MHz, 64-QAM modulation.}
\label{Fig:MIMO_BER_M1_PN}
\end{figure*}

We now study the system performance in the presence of phase noise. Fig. \ref{Fig:MIMO_BER_M1_PN} reports the system BER versus the transmitted power for the several considered transceiver architectures, for two values of the phase noise intensity, and for 4-QAM and 64-QAM modulations. In particular, we consider the cases of weak phase noise intensity, corresponding to the choice $L_{\rm PN}=-60$ dBc/Hz and $f_{\rm offset}=100$ Hz, and of strong phase noise, where $L_{\rm PN}=-89$ dBc/Hz and $f_{\rm offset}=1$ MHz (see the appendix for the definition of these parameters). Interestingly, we see that 64-QAM modulation, while being somewhat robust to weak phase noise,  is more sensitive than 4-QAM modulation to phase noise. This behavior is intuitively justified by noting that increasing the modulation cardinality the size of the Voronoi regions associated to the points of the modulation constellation gets reduced, thus increasing the likelihood that the phase noise leads to a wrong decision
on the transmitted symbol. 
As a consequence, the modulation cardinality should be chosen as a trade-off between the need to send as much as possible information per symbol interval, and the need to not increase too much the system BER, which, ultimately, leads to a reduction in the system throughput.

\subsubsection{Performance of adaptive learning algorithms} 
\begin{figure*}
\begin{center}
\includegraphics[scale=0.51]{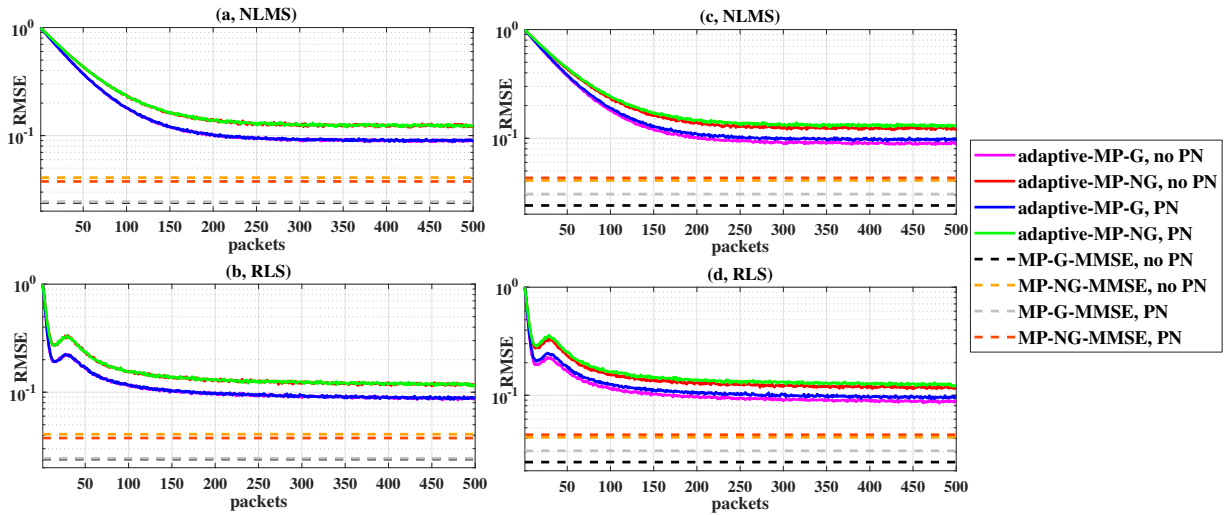}
\end{center}
\caption{Learning curve with $M=1$, $P_T=0$ dBW, 4-QAM modulation, $N_{\rm pkt}=500$, $L_T=200$ and CIB. In $(a)$ NLMS algorithm, $L_{\rm PN}=-60$ dBc/Hz and $f_{\rm offset}=100$ Hz; in $(b)$ RLS algorithm, $L_{\rm PN}=-60$ dBc/Hz and $f_{\rm offset}=100$ Hz; in $(c)$ NLMS algorithm, $L_{\rm PN}=-89$ dBc/Hz and $f_{\rm offset}=1$ MHz; in $(d)$ RLS algorithm, $L_{\rm PN}=-89$ dBc/Hz and $f_{\rm offset}=1$ MHz }
\label{Fig:Learning_4QAM_PN}
\end{figure*}

\begin{figure*}
\begin{center}
\includegraphics[scale=0.51]{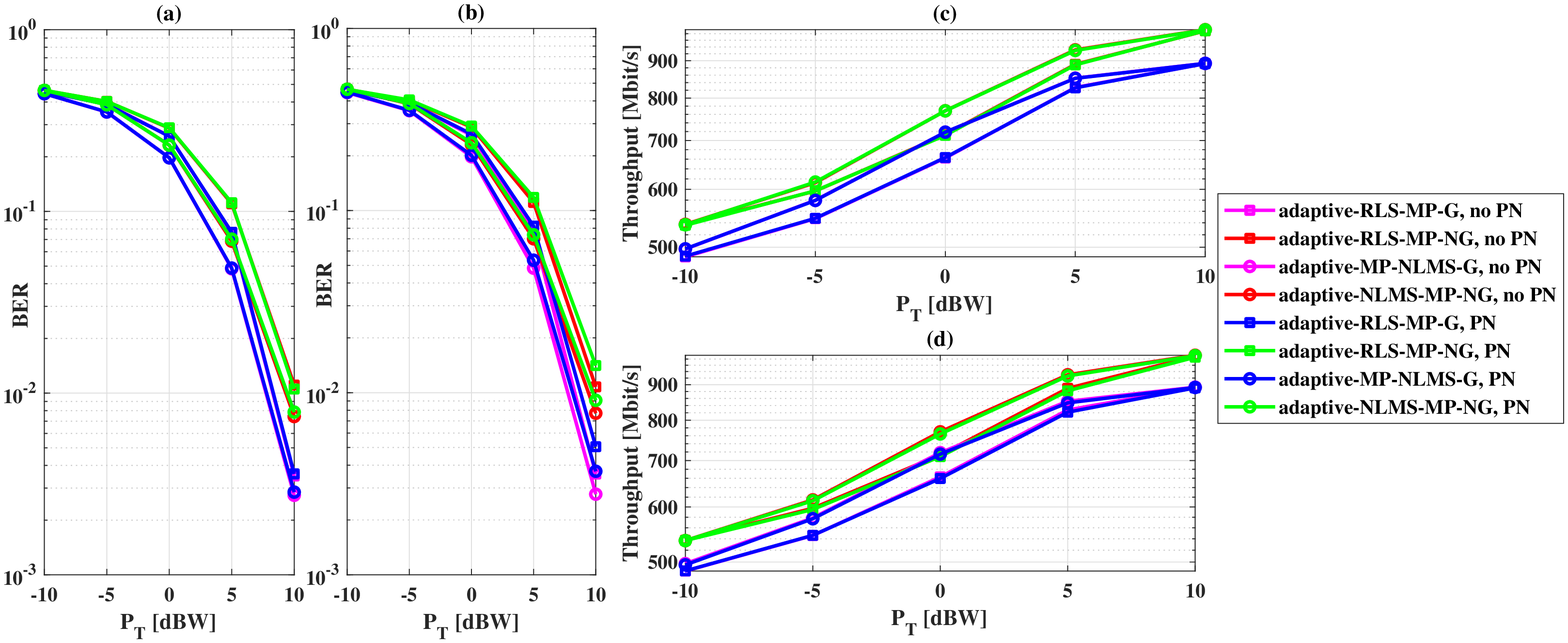}
\end{center}
\caption{BER and throughput performance of NLMS and RLS algorithms. In $(a)$ BER versus transmit power, $L_{\rm PN}=-60$ dBc/Hz and $f_{\rm offset}=100$ Hz; in $(b)$ BER versus transmit power, $L_{\rm PN}=-89$ dBc/Hz and $f_{\rm offset}=1$ MHz; in $(c)$ throughput versus transmit power, $L_{\rm PN}=-60$ dBc/Hz and $f_{\rm offset}=100$ Hz; in $(d)$ throughput versus transmit power, $L_{\rm PN}=-89$ dBc/Hz and $f_{\rm offset}=1$ MHz. }
\label{Fig:BER_througput_adaptive}
\end{figure*}

Finally, Figs. \ref{Fig:Learning_4QAM_PN} and \ref{Fig:BER_througput_adaptive} report the RMSE learning curves, and the steady-state BER and  throughput for the NMLS and RLS adaptive algorithms, respectively, for 4-QAM modulation. Again, two different intensities of the phase noise are considered, while both figures refer to the case of single stream transmission (i.e., $M=1$) and CIB is assumed. Fig. \ref{Fig:Learning_4QAM_PN} also report, as dashed horizontal lines, the steady-state RMSE achieved by the adaptive LMMSE algorithm performing the processing \eqref{UFMC_MMSE_estimated_reduced}. A comparison with the case in which no phase noise is present is also reported for benchmarking purposes.
Fig. \ref{Fig:Learning_4QAM_PN} shows that both the NLMS and the RLS algorithms exhibit a decreasing RMSE; it is shown that the recursive algorithms, as it is well-known, exhibit some excess error with respect to the adaptive LMMSE processing in \eqref{UFMC_MMSE_estimated_reduced}: this is indeed the price to be paid  to have lower computational complexity and enhanced tracking capabilities in time-varying environments. 

From Fig. \ref{Fig:BER_througput_adaptive} we can see that the recursive algorithms achieve good performance also in terms of steady-state BER and throughput. The gap between the curves corresponding to the presence of phase noise and the curves obtained in the ideal situation of no phase noise for the recursive algorithms is smaller than the gap observed for the LMMSE receiver \eqref{UFMC_MMSE_estimated_reduced}, even for the case of strong phase noise. The plots thus confirm that the proposed recursive algorithms, thanks to their tracking capabilities, have a higher degree of immunity to phase noise effects.


\section{Conclusions}
This paper has been focused on the design and the analysis of MIMO-UFMC transceivers operating at mmWave frequencies. Several signal processing schemes have been developed for data detection, taking into account the hybrid nature of the beamformers, and considering also the use of channel-independent beamformers at the transmitter. The proposed receivers have been extensively tested, also for the relevant scenario where multiple consecutive UFMC packets are transmitted with no inter-packet spacing. The results have shown that the proposed receivers are well capable of handling the increased interference arising from the absence of guard intervals, leading thus to an overall throughput increase of about 13 - 15 $\%$. The paper has also evaluated the effect of phase noise at the receiver; in particular, while using a modulation with large cardinality  leads to increased vulnerability to phase noise, results have shown that the proposed recursive implementations of the LMMSE algorithm exhibit a larger degree of immunity to phase noise. Overall, results have shown that UFMC is an outstanding modulation scheme that can be effectively coupled with MIMO architectures.  

The research results of this paper can be extended along many directions. First of all, beyond phase noise, other hardware imperfections such as non-linear power amplifiers might be considered and included in the analysis. Then, the coupling of the proposed transceivers with a massive MIMO architecture, wherein large antenna arrays are present at one side, if not at both sides \cite{buzzi2018energy} of the communication link, could be investigated. Finally, while the paper analyzed a single-user scenario, the extension of the proposed approach to a multiuser scenario is certainly worth being considered, along with the design of proper power control algorithms for sum-rate maximization.

\section*{Appendix - Phase noise modeling} \label{PN_Section}
In the following we provide details on the modeling of the phase noise at the receiver. 
Each RF chain at the receiver is assumed to contain a local oscillator (LO) with a synchronous distribution, i.e. a single reference signal is distributed to all the LOs in the RF chains which independently generate local oscillation signals\cite{petrovic2007effects,chen2017beamforming,puglielli2016phase}. 
Denoting by $\phi_i(n)$ the phase noise introduced by the LO in the $i$-th RF chain, we thus define the following $N_R^{\rm RF} \times N_R^{\rm RF}$ diagonal matrix containing the contribution of phase noise on each RF chain: \begin{equation}
\widetilde{\mathbf{\Phi}}(n)=\text{diag}\left( e^{j \phi_1(n)}, \ldots, e^{j \phi_{N_R^{\rm RF}}(n)}\right) \; .
\label{Phase_noise_RX}
\end{equation}
Given \eqref{Phase_noise_RX},  the output of the RF transceiver reported in Eq. \eqref{Y_BB}, in presence of phase noise at the receiver, can be represented through a matrix $\widetilde{\mathbf{Y}}_{BB}$ of dimension $[N_R^{\rm RF} \times (k+L+L_{ch}-2)]$, whose $n$-th column is expressed as
\begin{multline}
\widetilde{\mathbf{Y}}_{BB}(:,n)\!=\!\!\widetilde{\mathbf{\Phi}}(n)\!\!\! \sum_{\ell=0}^{L_{ch}-1} \sqrt{\frac{P_T}{M}} \mathbf{D}^H_{\rm RF} \widetilde{\mathbf{H}}(\ell) \mathbf{Q}_{\rm RF} \widetilde{\mathbf{X}}_{\rm BB}(:,n-\ell) \\ + \mathbf{D}_{\rm RF}^H \mathbf{w}(n) \; ,
\label{Y_BB_PN}
\end{multline}
with $n=1, 2, \ldots, k+L+L_{ch}-2$.
For the generation of the phase noise we use the procedure reported in \cite{Kasdin_Phase_Noise_Matlab}. In particular,  the sequence $\ldots, \phi_i(n), \phi_i(n+1), \ldots$ is obtained by considering a sequence of i.i.d. Gaussian real random variates and by passing them through a linear time-invariant filter. 

An IIR digital filter is used in which the numerator coefficient $\lambda_{\rm PN}$ is
\begin{equation}
\lambda_{\rm PN}=\sqrt{2 \pi f_{\rm offset}10^{\frac{L_{\rm PN}}{10}}} \; ,
\end{equation}
where $f_{\rm offset}$ is the frequency offset in Hz and $L_{\rm PN}$ is the phase noise level in dBc/Hz. The denominator coefficients $\gamma_m$ are recursively determined as 
\begin{equation}
\gamma_m=\left(m-2.5\right)\frac{\gamma_{m-1}}{m-1} \; ,
\end{equation}
where $\gamma_1=1$ and $m=1,\ldots, 64$.

\bibliography{finalRefs}
\bibliographystyle{ieeetran}

\end{document}